%% file: elsarticle-template.tex
\documentclass[5p,times]{elsarticle}

\usepackage{lineno,hyperref}
\usepackage{graphicx}
\modulolinenumbers[5]
\usepackage{xcolor, soul}
\usepackage{amsmath}

\newcommand{\rone}[1]{\textcolor{blue}{#1}}
\newcommand{\rtwo}[1]{\textcolor{red}{#1}}

\journal{Future Generation Computer Systems}

\begin{document}

\begin{frontmatter}

\title{Serverless data pipeline approaches for IoT data in fog and cloud computing}
\author[ut1]{Shivananda R Poojara}
\author[ut1]{Chinmaya~Kumar~Dehury}
\author[ut1]{Pelle Jakovits}
\author[ut2]{Satish Narayana Srirama\corref{mycorrespondingauthor}}

\cortext[mycorrespondingauthor]{Corresponding author}
\ead{satish.srirama@uohyd.ac.in}

\address[ut1]{Mobile \& Cloud Lab, Institute of Computer Science, University of Tartu, Tartu 50090, Estonia }
\address[ut2]{School of Computer and Information Sciences \\ \textit{University of Hyderabad,} Hyderabad 500 046, India }
\date{XXX}

\begin{abstract}
With the increasing number of Internet of Things (IoT) devices, massive amounts of raw data is being generated. The latency, cost, and other challenges in cloud-based IoT data processing have driven the adoption of Edge and Fog computing models, where some data processing tasks are moved closer to data sources. Properly dealing with the flow of such data requires building data pipelines, to control the complete life cycle of data streams from data acquisition at the data source, edge and fog processing, to Cloud side storage and analytics. Data analytics tasks need to be executed dynamically at different distances from the data sources and often on very heterogeneous hardware devices. This can be streamlined by the use of a Serverless (or FaaS) cloud computing model, where tasks are defined as virtual functions, which can be migrated from edge to cloud (and vice versa) and executed in an event-driven manner on data streams. In this work, we investigate the benefits of building Serverless data pipelines (SDP) for IoT data analytics and evaluate three different approaches for designing SDPs: \rone{1) Off-the-shelf data flow tool (DFT) based, 2) Object storage service (OSS) based }and 3) MQTT based. Further, we applied these strategies on three fog applications (Aeneas, PocketSphinx, and custom Video processing application) and evaluated the performance by comparing their processing time (computation time, network communication and disk access time), and resource utilization. \rone{Results show that DFT is unsuitable for compute-intensive applications such as video or image processing, whereas OSS is best suitable for this task. However, DFT is nicely fit for bandwidth-intensive applications due to the minimum use of network resources. On the other hand, MQTT-based SDP is observed with increase in CPU and Memory usage as the number of users rose, and experienced a drop in data units in the pipeline for PocketSphinx and custom video processing applications, however it performed well for Aeneas which had low size data units.}
\end{abstract}

\begin{keyword}
Serverless computing, data pipelines, cloud computing, fog computing, edge computing, internet of things
\end{keyword}

\end{frontmatter}

\input{Introduction}
\input{RelWorks}
\input{propopsedSDP}

\input{usecases}
\input{SDP_approaches}

\input{Experiment}

\section{Conclusions and future work} \label{sec:concFutWorks}
In this paper, we proposed three Serverless Data Pipeline (SDP) approaches: DFT, OSS, and MQTT-based SDP using Apache NiFi, MinIO, and MQTT services, respectively. We applied these approaches to three different fog computing applications namely Aeneas, Pocketsphinx and Video processing application. We investigated their performance using the metrics such as processing time \rtwo{ (computation time, disk access time and network access time)} and resource utilization (CPU, Memory, Network and Disk utilization) and rigorously analyzed the results by calculating a suitability index for each of them.  Results show that MQTT based SDP works best for Aeneas, DFT performs better for PocketSphinx and for video processing application, the OSS performance was good as compared with SDPs. However, an opportunity exist to improve the performance of proposed SDPs by scaling the serverless functions or using asynchronous invocations. Furthermore, the use of dynamic and stochastic workloads in future would help to evaluate the reliably, resilience and throughput etc, of the proposed SDP approaches.      



\section*{Acknowledgment}
This work is partially supported by the European Social Fund via IT Academy program and European Union's Horizon 2020 research and innovation project RADON (825040). We also thank financial support to UoH-IoE by MHRD, India (F11/9/2019-U3(A)).

\bibliography{mybibfile}

\end{document}

%% file: Introduction.tex
\section{Introduction}\label{sec:intro}

The advancements in internet technologies like 5G and other allied \rone{technologies have accelerated} the use of the Internet of Things (IoT) \cite{chang2019internet,toosi2019management} at a wider scale. With the increasing number of IoT devices, massive amounts of raw data is being generated. To cope with the ever increasing data, today's industries are looking towards management and computation solutions that enable to manage, process, and control the data flows in realtime, such as for Artificial Intelligence (AI) services.

 
To extract actionable insights from such raw data, it needs to be preprocessed, transported, transformed, and analyzed before useful knowledge can be extracted from it. To simplify designing and deploying such data processing services, pipelines are commonly used to compose a set of individual data processing processes into a single service, where the output of a component is the input of the next component. This allows to reuse and compose of common data handling processes into more complex data pipeline services~\cite{9127385}. 

The data pipelines would be deployed as software services running seamlessly in the Cloud or inside on-premise servers. \rone{ However, many IoT applications are event-driven and require performing actions in real-time \cite{aslanpour2021serverless}, which often requires large and expensive data processing clusters (e.g. Apache Spark, Flink, Storm) to be created to handle large-scale IoT data with self-adapting scaleable processing \cite{CARDELLINI2018171}}.

However, with the emergence of Serverless computing, a novel cloud computing service model that leverages the function level billing and scaling, designing event-based, real-time and scaleable IoT data processing has been significantly simplified \cite{baldini2017serverless}. Furthermore, such cloud-centric approach requires all data to be transported into one central data center and has multiple disadvantages, such as high dependency on end-to-end connectivity, higher latency, higher transfer and storage costs, and other typical issues with centralised data collection. 

This is where combining Serverless model and data pipelines can produce significant benefits to avoid some of the disadvantages of cloud-centric approach and reduce the complexity of designing multi-layer (Cloud, Edge, Fog) IoT applications.  
In data pipelines, each task is a process that consumes input data, manipulates it and produces output data and such tasks are composed into pipelines. This makes it easy to compose common data processing tasks into more complex data management services and potentially deploy some of the pipeline tasks closer to the data sources (e.g. Edge, Fog laters). 
In the Serverless model, functions are individually deployed services which are triggered on certain events (e.g. new database record or REST request arrival), receive data and produce output. Serverless cloud model has several benefits including fine grained auto scaling and increased productivity gains due to reusable serverless functions deployed either on on-premise or on the clouds \cite{casale2019radon}. It is also significantly easier to deploy individual functions in different locations compared to more monolithic applications (e.g. when compared to Apache Spark data analytic applications).

To combine both models, Serverless Data Pipelines (SDP) can be created when serverless functions are used as pipeline tasks and seamlessly invoked while the data moves through the pipeline. Serverless functions can be deployed in Cloud, Edge or Fog environments and data pipeline technologies are used for data transport, routing and function invocation. To provide one example: due to low latency demand and bandwidth constraints of cloud-centric approach, a novel fog computing architecture \cite{buyya2019fog,chang2019internet,article} was introduced between IoT devices and Cloud servers. Here, few of data analysis (data pipeline) tasks were moved from far-away clouds to remote fog nodes that are very near to data sources, this eventually improves the performance of real-time services. 

This approach is also helped by the advances in fog devices, which can now be equipped with enough compute (even with tensor processing units (TPU) and GPUs ) \cite{taherkordi2017iot} and storage capacities. It is very useful support to accommodate serverless frameworks that could accelerate the execution of data pipelines. However, the disparity of hardware resources available at the Edge, Fog and cloud leads to an interesting challenge of how to diversify and prioritize data flow between functions residing at fog and cloud to meet the expected  Quality of Service (QoS). There are also other related issues which need to be considered and evaluated. Serverless functions are stateless and its frameworks only deal with the runtime management of functions, completely separating it from the data management \cite{cheng2019fog}. This separation simplifies serverless computing but has drawbacks for data-intensive and stream processing pipelines \cite{hellerstein2018serverless} and that can lead to issue of dealing intermediate data between the functions in pipeline.

Considering the above mentioned issues and challenges, this work aims to investigate and compare the suitability of modern off-the-shelf Data Pipeline (DP) tools \cite{echo} \cite{truong2018integrated} and other frameworks (such as message brokers and object storage services) which are integrated with serverless frameworks and can be used to design dynamic Serverless Data Pipelines. 
We utilize three bandwidth and compute intensive real-time fog computing workloads: Aeneas \cite{mcchesney2019defog}, Pocketsphnix \cite{mcchesney2019defog} and a custom Video processing application to extensively measure the performance (w.r.t CPU, memory, disk and network usage) of different SDP architectures in the fog computing environment. 

\subsection{Motivation}

To process this data across edge/fog and cloud environments, a huge amount of resources (more than the actual demand) are allocated to process a user’s task in traditional systems, which is challenging and inevitable in resource-constrained edge/fog nodes. In this regard, serverless technology plays a significant role in the deployment of IoT applications by composing into stateless independent serverless functions across edge/fog and cloud environments. 

However, serverless functions are stateless with high granular scaling which introduces additional complexity and challenges in data management between functions residing at edge, fog, and cloud nodes. Some enterprise solutions such as Azure IoT or AWS Greengrass use serverless edge functions to pre-process and push data to enterprise clouds. Data movement between functions residing at the edge and cloud is often handled by using object storage services like AWS S3. However, it's challenging when a large set of functions are deployed in edge/fog infrastructure and data needs to be transferred on each function invocation. The object storage may yield higher charges when more data and more function invocations occur. Even-though, object storage attains the purpose of handling intermediate data but cost, latency, etc., are challenging. 

There also exist off-the-shelf DP tools like StreamSet and Apache NiFi, which provide some support for edge/fog environments and can also be utilized to solve the issues, but they usually manage the flow of data in a more centralized manner and often require significant computing resources to run effectively. 

Alternatively to object storage, its also possible to use data brokers (e.g. Apache Kafka, MQTT) as Message Queues between serverless functions for designing serverless data pipelines. Compared to object-storage, they would require less storage and may be faster due to more extensive memory usage, which is highly desirable in edge/fog environments. However, compared to NiFi, it may be more difficult to control the precise execution flow of pipelines. 

The aforementioned challenges have motivated us to investigate the advantages and disadvantages of different mechanisms for integrating serverless platform with data pipeline platforms. The following subsection will list the contributions in the proposed work.
 
\subsection{Contributions}
In the above context, our contributions in this work can be summarized as follows:
\begin{itemize}
	\item We demonstrate how SDP can be deployed in three layered IoT architectures. 
	\item We propose three approaches for designing Serverless Data Pipelines with different data handling mechanisms (Apache Nifi, Message Queues and object storage services such as AWS S3). 
	\item We use real time fog computing workloads such as Aeneas, PocketSphinx and custom Video processing applications to compare the performance (such as processing time) and resource utilization of these different SDP approaches.
	\item We provide insights on the suitability of these SDPs for different types of fog computing workloads.  
\end{itemize}
 
The rest of the paper is organized as follows. In section~\ref{sec:relWorks}, we present literature survey on current state of art data pipeline technologies and ecosystem. \rone{The proposed SDP architecture and real time IoT usecases are described in the section~\ref{sec:probForm} and \ref{sec:usecases}, respectively. Following this,  three novel SDP approaches are designed and articulated and implemented for real time fog computing use cases in section~\ref{sec:sdpApproaches} and compared with different performance metrics in section~\ref{sec:expsetup}}. Finally, the concluding remarks and the future works are discussed in section \ref{sec:concFutWorks}.

%% file: RelWorks.tex
\section{Related works}\label{sec:relWorks}
A number of SDP architectures and solutions have been proposed in the field of IoT data management in edge, fog and cloud environments. This section briefly summarizes the recent work done in the context of SDP architectures and models.



The public cloud service providers such as \textit{AWS greengrass} \cite{AWS}, \textit{Google Cloud IoT} \cite{google}  and Microsoft- \textit{Azure IoT Edge} \cite{Azure} have typical IoT data pipeline solutions for industrial, healthcare, smart city and other real time use cases. For example, consider AWS IoT Greengrass, where Lambda service will be executed at the edge layer for data acquisition and pre-processing. Later data is forwarded to the cloud by edge devices and then it passes through pipeline of activities for post processing and finally is delivered to the data sink. 

\rone{Valeria et al. proposed a solution of IoT data stream processing in distributed fog and edge computing environments with decentralized scaleable manner \cite{CARDELLINI2018171} and further extended to how data processing operators were placed in computing nodes considering the efficiency, application topology and resources configurations \cite{8630099}. These works provide hint that off-the-shelf data stream processing tools such as Apache Storm can be used for the task. However, these stream processing tools require huge computing clusters and in IoT deployments more often devices are heterogeneous with limited computing capacity. More often IoT workloads are event and time driven which motivates us to investigate the serverless based data processing pipelines. Further, SDPs easily been deployed at various levels in the IoT hierarchy (Edge, Fog and Cloud Infrastructure) with efficient granular scaling of the serverless functions. }

Das et al. \cite{9139674} proposed a model for efficient execution of user tasks as serverless functions in edge/cloud environments and designed a set of data pipelines using AWS Greengrass on edge devices along with Lambda capabilities. Our approach looks similar to this model, however it lacks fog based processing pipeline model.

Dehury et al. \cite{dehury2020ccodamic} designed a framework known as CCoDaMiC, which aims to ensure data accuracy, trustworthiness, and validation in SDP. This work directly relates to our proposed DFT based SDP approach. However, CCoDaMiC mainly focuses on data accuracy and trustworthiness and not on the performance of the applications. Lixiang et al. \cite{ao2018sprocket} designed a framework for video processing using serverless lambda functions known as \textit{Sprocket}. Authors demonstrated the efficiency of serverless functions for faster execution by constructing pipeline of activities for video handling. However, their primary focus is to reduce latency and cost by using the  techniques of  parallelism. Interestingly, this work motivated us to consider the complex video processing use case in our proposed research. 


Several techniques and methods have been proposed illustrating the use of MQTT for data acquisition from different data sources via publish/subscribe model \cite{helu2020scalable,akin2020enabling,ronkainen2015designing}. MQTT brokers can act as data carriers and can store data until subscribers consume it. This approach is well suited to store temporary or intermediate data between processing elements in SDP. In our work, one of the approaches uses MQTT together with serverless framework to construct data pipelines from data source to sink.

Thus to the best of our knowledge, none of the research works attempted to investigate and compare the different techniques in the construction of SDP with different approaches for intermediate data handling between serverless functions.  


%% file: propopsedSDP.tex
\section{Proposed System} \label{sec:probForm}

\begin{figure}[ht]
    \centering
    \includegraphics[width=0.95\linewidth]{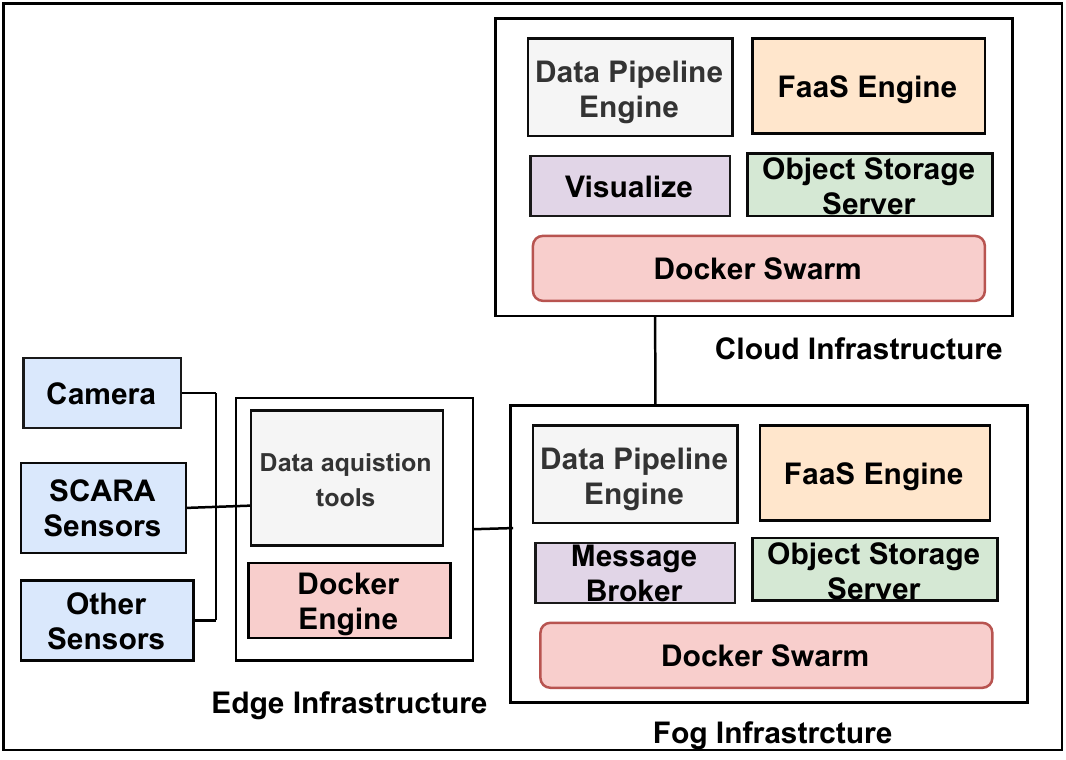}
    \caption{Proposed SDP architecture}
    \label{fig:f1}
\end{figure}

Based on the literature survey, we designed an overall SDP architecture and the required software services to handle the flow and execution of data in a pipelined manner, as shown in Figure~\ref{fig:f1}. The data are generated by the IoT devices (such as surveillance cameras, SCARA (Selective Compliance Assembly Robot Arm) robot sensors, health monitoring sensors etc.) and are eventually sent to cloud infrastructure for processing and storage.

Instead of sending the data directly from IoT device to the cloud, the data is preprocessed by different computing environments such as edge computing devices or fog servers through the execution of serverless functions. As in Figure \ref{fig:f1}, the data are sent to Edge infrastructure, followed by the Fog infrastructure. Both the edge gateways and the fog infrastructure are responsible for providing the infrastructure to host data management and serverless frameworks. 


\subsection{Edge Infrastructure}\label{sec:sysarch:edgeInfra}
Edge infrastructure is mainly responsible for receiving the data from IoT devices. For this purpose, different data acquisition tools or software solutions can be used such as MiNiFi or SDC Edge, custom services such as Python or other run-time services, as shown in Figure \ref{fig:f1}. These solutions usually come with limited capabilities to process the data. Upon receiving and processing, edge infrastructure forwards the data to fog infrastructure. The detailed description of the services used in edge infrastructure are described below.
\begin{itemize}
  \item \textit{MiniFi~\footnote{https://nifi.apache.org/minifi/}}: Apache MiNiFi is a super light-weight version of NiFi made for the edge devices. It can run as a system service, and it is centrally managed using Apache NiFi. Developers can easily design the pipelines using a set of processors in Apache NiFi and push them into the MiniFi service. These pipelines can handle preliminary data operations near to the source, e.g., compressing a video recorded by drone before sending to cloud/fog to reduce bandwidth consumption, etc. We use this MiniFi service in the implementation of DFT tool based SDP as described in \ref{subsec:Queueless}.
  \item \textit{Custom services}: Custom services are similar to the MiniFi processors. Such services need to be created from scratch using a specific programming language. E.g., A Python program can be created to collect and compress the video. Python-based custom services are created and used in the implementation of OSS  and MQTT-based SDP as described in \ref{subsec:perQueu} and \ref{subsec:mqtt}, respectively. 
\end{itemize}

\subsection{Fog Infrastructure} 
Fog Infrastructure is mainly responsible for processing the data received from Edge infrastructure, for which, Data pipeline engine (Apache NiFi\footnote{https://nifi.apache.org/}), FaaS engine (OpenFaaS\footnote{https://www.openfaas.com/}), MinIO\footnote{https://min.io/}, and MQTT\footnote{https://mqtt.org/} services are used. The fog infrastructure includes a group of fog servers deployed in a cluster with a set of particular software services using Docker Container Engine. The processed data are then forwarded to the cloud infrastructure to further process, store, and generate alerts and notifications. The use and necessity of software services are described below: 

\textit{FaaS Engine}: FaaS Engine is primarily one of the core components of this proposed work. Several open-source serverless platforms are available such as Apache OpenWhisk, OpenFaas, Kubeless, etc. OpenFaaS serverless platform is used in this work, as it is lightweight and easy-to-configure over other alternative solutions. OpenFaaS can be installed atop of Docker or Kubernetes platform. Docker containers are used to host and execute the serverless functions. The functions can be invoked using HTTP endpoints with the necessary data. The function invocation is performed in the pipeline by Data pipeline engine, MinIO event notification system, and MQTT event notification service.
    
\textit{Data pipeline engine}: As discussed in Section~\ref{sec:intro}, we use Apache NiFi, a data pipeline processing platform, which manages the data flow between the systems. This provides a set of independent processors with specific functionalities to process and manage the data. The data flow between processors is managed via scalable queues. Developers can easily design custom data pipelines using a flexible user interface and automatically configure, control, and deploy the pipelines in Edge infrastructure using MiniFi service. This MiniFi-Nifi integration efficiently manages the orchestrated IoT data processing from the edge, fog, and cloud, and vice-versa.
   
\textit{Message Queue}: Message Queues are published/subscribe protocol based data carriers between source and sink. We demonstrate the use of Message Queues to build data pipelines integrated with the OpenFaaS serverless platform. A light-weight messaging protocol, MQTT, is ideal for small sensors and mobile devices and is suitable for high-latency or unreliable networks. MQTT uses different data types such as UTF-8 encoded string, bit/byte integer, binary data, and UTF-8 string pair. A serverless function can publish the processed data to MQTT, and in turn, functions are invoked when data need to be subscribed using web-hooks. The flow of data between MQTT and serverless platform builds consistent, reliable SDP.
    
\textit{Object storage service}: An open-source cloud-based storage solution, MinIO, compatible with Amazon S3, stores the IoT data. This provides a RESTful API to access/insert/remove buckets and objects. Moreover, triggers are set to bucket when its content is accessed/written/removed, and corresponding event notifications are generated using techniques such as web-hooks, Message Queues. This is advantageous in IoT applications for handling event-driven data. It is configured as high-availability cluster using docker swarm.

\subsection{Cloud Infrastructure}
Cloud infrastructure is mainly responsible for processing heavy computation data received from Fog infrastructure and storing the data. This is also responsible for generating alerts and notifications to activate other business processes whenever required. To perform this, a set of services from cloud providers or user-configured open-source services are used, such as Data pipeline engine (Apache NiFi, AWS data pipeline, Google Cloud pipeline, etc.), Object storage service (MinIO/AWS S3/ Google object storage), Message queues (MQTT/AWS SNS), and Faas engine (OpenFaaS/AWS Lambda/Google Functions).

The setup and configuration of the FaaS engine, data pipeline engine, and object storage service are the same as that of the fog infrastructure. Along with this, the cloud infrastructure is also responsible for Visualization and Reporting. The primary job of the visualization is to display processed data using visualization tools. The Grafana visualization tool is used to measure Edge/Fog and cloud nodes' performance metrics in this work. Additionally, the Prometheus time-series database is used where performance metrics are collected.

%% file: usecases.tex
\section{Realtime IoT Use Cases} 
\label{sec:usecases}
To compare and evaluate the performance of proposed  SDP approaches, a set of \rone{standard  fog computing workloads or applications are} considered from the article \cite{mcchesney2019defog}. The applications are categorized into  latency critical (LC), bandwidth intensive (BI), location aware (LA) and computational intensive (CI). The corresponding applications from the article \cite{mcchesney2019defog} such as Aeneas (BI), PocketSphinx (BI, CI), Yolo object detection (BI, CI) were considered as real time IoT workloads. These applications are redesigned into sequence of data flow pipelines that can spread across edge, fog and cloud infrastructure.  The detail implementation of data flow pipelines are explained below.

\subsection{\textbf{Custom video processing using Deep learning based object classification using YOLOv3 and ffmpeg tool}} The You Only Look Once (YOLO) makes predictions with a single network evaluation unlike systems like Region-based Convolutional Neural Networks (R-CNN) which require thousands of networks for a single image. Hence, its predictions are 1000x faster than R-CNN and 100x faster than Fast R-CNN. 

This fog application is ideal candidate to consider because of bandwidth sensitiveness and requires high network bandwidth to send video stream from edge node to cloud and aiming for faster processing with immediate response. Most of the operations are compute intensive and demand for more CPU and Memory resources.  \rone{So, the application} is designed to process preliminary object detection in fog and send to cloud for further analysis.
\begin{figure}
    \centering
    \includegraphics[width=0.95\linewidth]{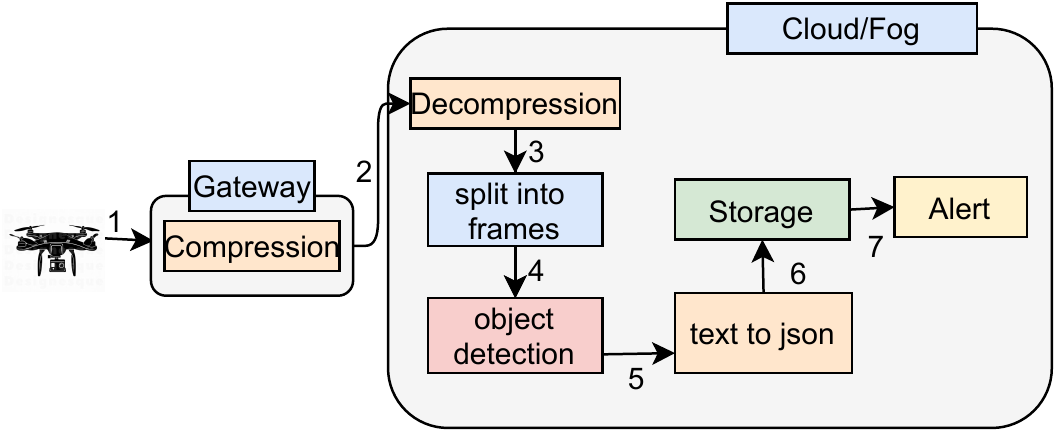}
    \caption{Abstract view of video processing  data pipeline}
    \label{fig:yolo}
\end{figure}
To understand precisely, we use a scenario of real time object detection from the article \cite{renart2018edge}.  The abstract flow of data in this use case is shown in Figure \ref{fig:yolo}. Here a drone is used to capture the video footage and to process this, sequences of operations are carried out as follows:
\begin{enumerate}
    \item Drone captures video footage and forwards to edge gateway using communication protocol such as MQTT or HTTP or other protocols.
    \item Edge gateway compress the video and forwards to fog nodes.
    \item In a fog node, video is decompressed using set of tools (gzip or zip).
    \item The video will be split in to number of frames that can be processed individually.
    \item The frames are passed to YOLOv3 framework and objects are identified from respective frame.
    \item The raw output generated from YOLOv3 is processed to json document. This will nicely arrange the raw text into json elements consisting of identified objects. 
    \item The json documents are stored in storage service for further analysis.  
    
\end{enumerate}

To design the above example in traditional computing platforms, the developer needs to specify a necessary input, configuration, and run-time environment to perform required data operation seamlessly by provisioning resources on the fly. However, it adds an issue of over-provisioning than demand. For example, short-running tasks like triggering an alert message to the end-user when a human/animal object is identified don't require heavy computation. Thankfully, the Serverless platform can subsidize this issue by invoking functions whenever events are triggered by consuming less computation resource. 
In this application,  entire video processing application is decoupled into a set of OpenFaaS-based serverless functions as shown in Table~\ref{table:functions}. Among which, two major functions are (a) \textit{Split} and (b) \textit{Yolo} as described below.
\begin{itemize}
    \item \textit{Split}: Splits the video into multiple frames using ffmpeg, an image/video editing tool. The number of splits depends on the value given to the fps argument in ffmpeg command
    \item \textit{Yolo}: Yolo is a object detection framework that has a darknet library. 
\end{itemize}
The rest of the functions mentioned in Table~\ref{table:functions} are used as subsidiary functions in the pipeline. This video processing application is implemented in all the three SDP approaches, as described in below Section~\ref{sec:sdpApproaches}. 

\begin{table} [h!]
\centering
\caption{Number of serverless functions used}
\label {table:functions}
\footnotesize
\begin{tabular}{ | c c c c | }
\hline
SDP/Application & Aeneas & Pocketsphinx & Video proc.   \\ 
\hline \hline
DFT based SDP & 2 & 3 & 3 \\
OSS based SDP & 5 & 6 & 6\\
MQTT based SDP & 2 & 3 & 3 \\ 
\hline
\end{tabular}

\end{table}


\subsection{\textbf{Aeneas: A text-audio synchronisation }} The Aeneas tool is specialized for automated synchronization of audio to given text file also known as forced alignment. It automatically generates a synchronization map between a list of given text fragments and an audio file containing the narration of the text. 
\begin{figure}
    \centering
    \includegraphics[width=0.95\linewidth]{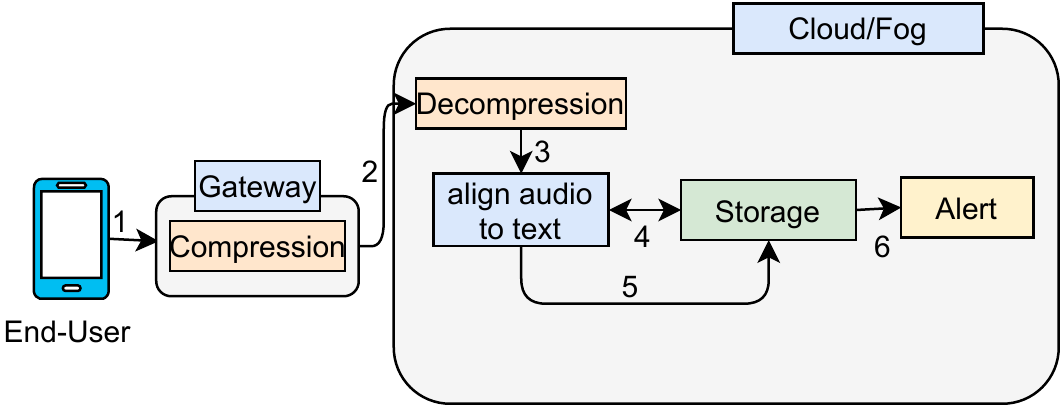}
    \caption{Abstract view of Aeneas data pipeline}
    \label{fig:aeneas}
\end{figure}

This fog application is ideal candidate to consider because of bandwidth intensive and consumes huge network bandwidth to stream audio files from large set of end user devices to edge node and cloud node. End users aim for faster response times and to achieve this, few of the operations are performed in fog node. We show the scenario of Aeneas based real time application and its abstract flow in Figure~\ref{fig:aeneas}.

Here, end-user device such as mobile device or other devices are used to stream the .wav or .mp3 audio files. Then the following operations are carried out as follows:
\begin{enumerate}
    \item End user offloads a .wav file to edge gateway using mobile device.
    \item Edge gateway compress the audio and forwards to fog nodes.
    \item In a fog node, audio is decompressed using set of tools (gzip or zip).
    \item The file (.xhtml) is downloaded from the cloud or other repository used as a input to Aeneas tool. The file contains a text used for alignment in the audio file.
     \item The raw .mp3 or .wav is processed using Aeneas tool along with the given file (.xhtml). 
     The Aeneas tool has facility to generate the alignment output in json.
    \item The json documents are stored in storage service for further analysis.  
    
\end{enumerate}


The above traditional data pipeline is redesigned to SDP by composing these operations in to serverless functions.   It has one major function:
\begin{itemize}
    \item \textit{aeneas}: Aeneas is python library for forced alignment of given text in a audio file and is configured with python3 run time. This function accepts two input files text file (.txt, .xhtml), audio file (.wav,.mp3) and generates the aligned output (.json, .mile). It generates .json as a HTTP response to the invocation.
    
\end{itemize}



\subsection{\textbf{PocketSphinx: A Speech-to-text conversion} }
Its a software engine specialized for speaker-independent continuous speech recognition \cite{mcchesney2019defog}. An audio file (.wav)  is converted to a defined language in text form using a pre-trained acoustic model to determine the source and destination language for speech-to-text conversion. The sample audio files are taken from large scale speech repository \footnote{http://www.repository.voxforge1.org/downloads/SpeechCorpus}.

\begin{figure}
    \centering
    \includegraphics[width=0.95\linewidth]{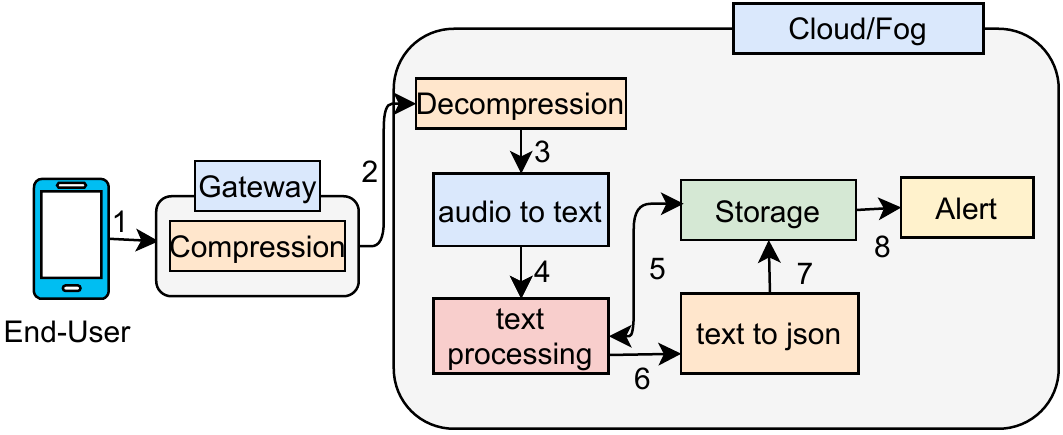}
    \caption{Pocketsphinx application data pipeline}
    \label{fig:pocketsphinx}
\end{figure}

In this fog application, we consider a scenario, where end user submits a .wav file via mobile phone and it needs to be processed (either fog/cloud) to find a given text in the audio file. This application is bandwidth and compute intensive, and requires higher network bandwidth to offload the audio files to cloud. Hence it is feasible to process near to the data source in the fog nodes to achieve higher response times. The Figure~\ref{fig:pocketsphinx} shows the abstract view of the application with following set of operations: 

\begin{enumerate}
    \item End user offloads a .wav file to edge gateway node
    \item It will be compressed  and forwarded to fog/cloud infrastructure
    \item Decompress the audio file
    \item Process an audio file to text format using PocketSphinx tool
    \item Get the text to search in output produced from PocketSphinx operation
    \item Perform text search operation 
    \item Convert into proper json document
    \item Store into storage service for further analysis or usage and send corresponding alerts.
\end{enumerate} 

The overall flow of the above mentioned PocketSphinx application is designed in to set of serverless functions as shown in Table~\ref{table:functions}. It has one major function:
\begin{itemize}
    \item \textit{pocketsphinx}: Pocketsphinx is python library for speech-to-text conversion using predefined acoustic model. This function accepts one  audio file (.wav,.mp3) and generates the aligned output (.json). It generates .json as a HTTP response to the invocation.
This PocketSphinx application is implemented in all the three SDP approaches, as described in the Section~\ref{sec:sdpApproaches}.    
\end{itemize}

\rone{ We represent $F = \{f_1,f_2,\dots,f_m\}$ as a set of serverless functions for each application, for example Aeneas for DFT based SDP has $m=2$. Further in the below section, we describe the specific implementation of SDP approaches and associated three usecases that are implemented according to design of proposed SDPs. }



%% file: SDP_approaches.tex
\section{Serverless Data Pipeline (SDP) approaches} \label{sec:sdpApproaches}
Considering the above overall architecture and challenges mentioned in motivation section, the serverless data pipeline can be designed by following different approaches. In this section, we introduce three approaches: (a) Off-the-shelf data flow tool based SDP, (b) Object storage service based SDP, and (c) MQTT based SDP. Further, we implemented the proposed SDPs for real time IoT usecases as described  in the Section~\ref{sec:usecases}. 

\subsection{Off-the-shelf data flow tool (DFT) based SDP }\label{subsec:Queueless}
In DFT based SDP, the developer need not have to maintain the queue, rather we use the queuing capability of Apache Nifi, which manages the centralized management of data and with integration of serverless platform. 

\begin{figure*}[!ht]
 \includegraphics[width=0.95\linewidth]{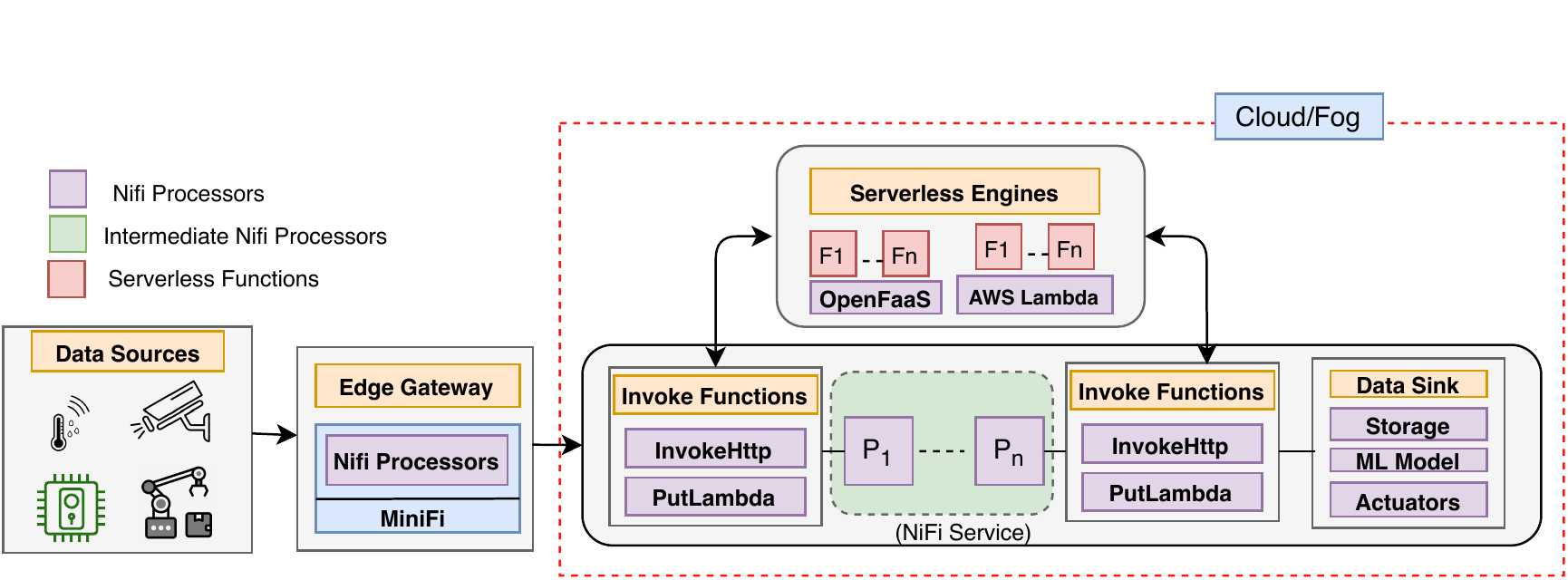}
 \caption{DFT based SDP approach using Apache Nifi and OpenFaas.}
 \label{fig:f2}
\end{figure*}

In this SDP approach, we have used Apache Minifi in the Edge infrastructure to receive and preprocess the data, as shown in Figure \ref{fig:f2}. As discussed in Subsection~\ref{sec:sysarch:edgeInfra}, MiniFi service can run on resource-constrained devices and is autonomously managed using central Apache NiFi from Fog Infrastructure. Here, sensed data from IoT devices is received into MiniFi using processors such as \textit{ConsumeMQTT} or \textit{ListenHTTP}, based on the communication protocol used between edge infrastructure and IoT devices. Some data preprocessing processors such as data compression, filtering, and aggregation are also configured but with limited computing resources and capabilities. MiniFi does not have GUI and is developed with java/C++ libraries and can quickly be started as a system service. The data flow with processor groups is designed in Apache NiFi and is automatically pushed into the MiniFi service. MiniFi performs the data operations and pushes the flow file containing the data to the Apache NiFi service configured in the fog. 

Apache NiFi is used to handle data flow in fog and cloud infrastructure. 
Apache NiFi provides a flexible set of processors for data operations and integration between cross platform systems. This gives the capabilities to seamless integration of serverless platform through specific Nifi processors, such as \textit{InvokeHttp}, \textit{PutLambda}, etc. Such multiple NiFi processors can be connected to others, allowing the developer to invoke multiple serverless functions. The key benefit is that Apache NiFi facilitates queued data that can lend back pressure when limits are attained during data flow processing. Another benefit is that it has priority queuing to set single/multiple prioritization schemes that dictate how data is retrieved from a queue. 
This allows the developer not to pay much attention to maintain or implement the queue between each pair of serverless function invocations. 

For the serverless platform, OpenFaaS is used in both fog and cloud infrastructures. However, public cloud serverless services can also be used in the cloud infrastructure such as Amazon Lambda. The serverless function receives the data flow file as input from \textit{invokedHTTP} request and sends the processed data as an HTTP response body. Every serverless function is invoked using an HTTP endpoint with respective HTTP methods (POST or GET).  \rone{We implemented the DFT based SDP using real time IoT use cases as described in the below paragraphs.}

\rone{In \textbf{Custom video processing application}, the Drone sends the video footage to the edge node. In the edge node, MiniFi service is configured with \textit{GetFile} MiNiFi processor to read video file from disk and forward the video to Nifi in the fog node. The Nifi in fog infrastructure consists of three main processors to invoke three different OpenFaaS serverless functions (FFmpeg, Yolo, convertTojson). On the other hand, the Nifi in Cloud infrastructure consists of multiple processors to store the data (received from Fog infrastructure) into MinIO bucket.}

\rone{Similarly, in the  \textbf{Aeneas} application the end-user mobile device sends an audio file to the edge node. In the edge node, MiniFi service is configured with \textit{GetFile} MiniFi processor to read audio file received and compressed using gzip processor and forward it to NiFi configured in the fog node. The NiFi in fog infrastructure consists of three main processors, first is  decompress processor, second is to invoke two different OpenFaaS serverless functions (aeneas, getFile(.xhtml)), third is creating JSON from the output. Finally, this data is stored in the MinIO bucket using multiple NiFi processors.}

\rone{The \textbf{PocketSphinx} has similar implementation as above, but Nifi in fog infrastructure consists of three main processors; first is  decompress processor, second is to invoke two different OpenFaaS serverless functions (pocketsphinx, text-processing), third is creating json from the output.}

\subsection{ Object Storage service based SDP}\label{subsec:perQueu}
In OSS based SDP, the object storage service will resemble a persistent queue where the developer can visualize the data stored in the storage server as a queue. The capability of object storage service is to trigger an event notification using web-hooks, making the integration of object storage with serverless platform flexible. 

\begin{figure*}[!htb]
 \includegraphics[width=0.95\linewidth]{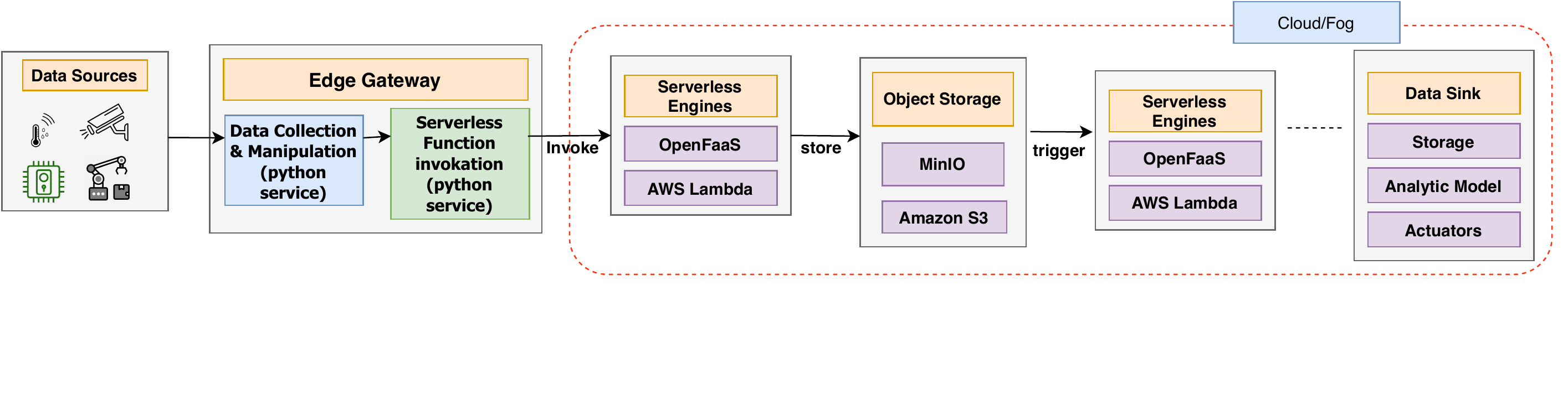}
 \vspace{-12mm}
 \caption{OSS based SDP approach using MinIO. }
 \label{fig:f3}
\end{figure*}

In this approach, we have used Python service in edge infrastructure to receive, preprocess, and transport data from IoT devices to fog infrastructure, as shown in Figure~\ref{fig:f3}. Here, Python \textit{requests} library is used to invoke serverless functions from edge Python service to functions residing in fog.

MinIO is used to handle data flow in fog and cloud infrastructure. MinIO is an open-source high-performance scalable storage service, as described in section \ref{sec:probForm}. The flexible bucket notifications are a set of events such as inserted, accessed, deleted and copied. The corresponding events are triggered using Web-hooks. This flexibility makes seamless integration of storage service and serverless platform invocations. Apart from this, the MinIO Python client library makes it easy to code the functions to access the object data from a specific bucket.

For the serverless platform, OpenFaas is used in both fog and cloud infrastructure. Here, the serverless functions are invoked from the gateway node with data or from the events triggered in MinIO buckets. Furthermore, the serverless function may store processed data into buckets using the MinIO client library. Again, events may trigger to invoke functions and continue until the fog node forwards data to cloud infrastructure. In the cloud infrastructure, data flows in a similar fashion over MinIO and OpenFaas serverless platform. This constitutes an SDP, where object storage with persistent mode acts as an intermediate data handling mechanism between serverless functions. \rone{ The following paragraph will describe the use of OSS based SDP in designing the real time use cases. }

\rone{In \textbf{Custom video application}, we use a Python service as a drone simulator to send the video file to the gateway node. The gateway node is configured with a Python service to read a video file and send it to the fog node for processing. In fog node, MinIO is configured with two buckets: (a) \textit{unprocessed} to store raw images and (b) \textit{processed} to store processed video in JSON format. These buckets are set with web-hook event notifications to trigger serverless functions when a new data object is inserted. This implementation uses five serverless functions, as given in Table \ref{table:functions}. However, in \textbf{Aeneas application}, in fog node, MinIO is configured with two buckets: (a) \textit{raw-audio} to store raw images and (b) \textit{syncmap} to store synchronization map generated  to audio file in JSON format. This implementation uses six serverless functions, as given in Table \ref{table:functions}. Finally,  the \textbf{PocketSphnix application}  in the fog node, MinIO is configured with two buckets: (a) \textit{raw-pocketsphinx} to store raw images and (b) \textit{processed-pocketsphinx} to store the converted audio file to text, (c) \textit{output-pocketsphinx} to store text-processed data.  This implementation uses 6 serverless functions, as given in Table \ref{table:functions}. In the cloud infrastructure, two MinIO buckets are created, (a) \textit{success-pocketsphinx}- to store success results from text processing, (b) \textit{failure-pocketsphinx} to store failure results from text processing. These buckets are responsible to store the processed audio files and output is mainly in text format.}

\subsection{ MQTT-based SDP}\label{subsec:mqtt}

\begin{figure*}[ht]
 \includegraphics[width=0.95\linewidth]{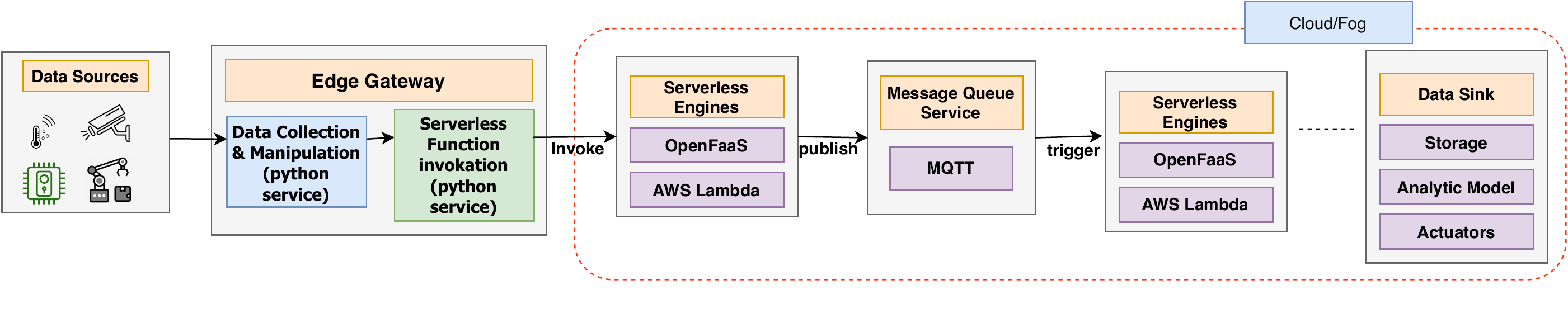}
 \caption{MQTT based SDP Architecture}
 \label{fig:f4}
\end{figure*}

In this proposed SDP approach, MQTT is used as a queue to store the data, different queues are represented by different MQTT topics and serverless functions can be triggered when new data objects are published into a specific topic. The gateway node receives data from IoT device using the custom Python service (Python code is written to perform specific operation). The received data are preprocessed and published to the MQTT broker with a topic name, as shown in Figure~\ref{fig:f4}. 

The topic names need to be subscribed by the OpenFaaS serverless functions, to consume the data in the queue. For this, Serverless frameworks should be built in with connectors between itself and message broker to subscribe the topics and invoke corresponding functions. Apart from this, MQTT doesn't have the capability to directly trigger an HTTP endpoint upon the arrival of new data. Thanks to the OpenFaaS community for developing the \textit{openFaas-mqtt} connector, which runs as a service to invoke serverless functions by subscribing to MQTT topics. OpenFaas has multiple connectors supported for different Message Queues. The serverless function processes the data and publishes the output again to the Message Queue. This process continues until the data from the source reaches to the data sink, as shown in Figure~\ref{fig:f4}. \rone{The following paragraphs describe how the use cases were implemented using the MQTT based SDP. }

\rone{Similar to OSS based SDP approach, all the three applications are configured with python service in the edge node. In \textbf{Custom video processing application}, the Python service publishes the video to MQTT broker with topic name. The \textit{openfaas-mqtt} connector running in fog node subscribes to the topic name and invokes the functions. Here, we use three serverless functions, as given in Table \ref{table:functions} and one \textit{openfaas-mqtt} connector service. Similarly, in the \textbf{Aeneas application}  the Python service publishes the audio file to MQTT broker with topic name. The \textit{openfaas-mqtt} connector running in fog node subscribes to the topic name and invokes the functions. Here, we use three serverless functions, as given in Table \ref{table:functions} and one \textit{openfaas-mqtt} connector service. 
In \textbf{PocketSphinx application}, we use three serverless functions, as given in Table \ref{table:functions} and one \textit{openfaas-mqtt} connector service.} 

\rone{For calculating the evaluation metrics in Section \ref{sec:expsetup}, we represent $S =  \{S_1,S_2,\dots,S_k\}$ as set of storage units. The storage unit could be processors in NiFi, MQTT queues or MinIO buckets. Next section will describe about experiment details.}


%% file: Experiment.tex
\section{Experiment and results} \label{sec:expsetup}
All the  proposed SDP approaches are implemented on three applications as described in the  Section~\ref{sec:usecases}. Further, the goal is to measure the performance w.r.t metrics to understand and investigate  efficiency of those SDPs on various applications (text, audio, video and image applications).   In the following section, we will discuss the  metrics used to measure the performance and analysed the results, further outlined  the experience on the SDP implementation and provided future directions.

\subsection{Performance metrics}
\begin{table}[t]
\caption{\rtwo{List of Notation.}}
\begin{tabular}{|p{0.15\linewidth}|p{0.75\linewidth}|}
    \hline
    \textbf{Notation} & \textbf{Description} \\ \hline
    $P$ & Computation time    \\ \hline
    $C_T$ & Communication time    \\ \hline
    $F$ & Set of serverless functions\\ \hline
    $R$ &  Set of $n$ number of concurrent users' requests $R = \{ r_1, r_2, \dots, r_n\}$ \\ \hline
    $D$ & Duration of the user request from source to destination (data sink)  \\ \hline
    $D_{at}$ & Timestamp recorded at arrival from the source   \\ \hline
    $D_{ct}$ & Timestamp recorded at destination    \\ \hline
    $DAT$ & Disk Access Time   \\ \hline
    $NCT$ & Network Communication Time   \\ \hline
    $S$ & Set of intermediate storage units $S=\{S_1,S_2,\dots,S_k\}$  \\ \hline
     $DU_{at}$ & Timestamp recorded when data unit arrived in the storage unit   \\ \hline
     $DU_{dt}$ & Timestamp recorded when data unit departed from the storage unit   \\ \hline
\end{tabular}
\label{table:Notation}
\end{table}

\rtwo{In this subsection, we will describe  various performance metrics along with their mathematical formulae. The resource utilization metrics are measured cumulatively on all the three layers of infrastructure  that consists of utilization of edge (Gateway) resources, fog resources and cloud infrastructure resource. Here, concurrent user requests are generated in the sensor nodes in all the three use cases with corresponding data such that it mimics the real time application. The \textit{Prometheus} and \textit{Node Exporter} software solutions are used to collect such metrics. \textit{PromoQL} (Prometheus Query Language) is used to calculate the metrics for specific time period.}
\begin{itemize}
    
    \item \rtwo{\textbf{Processing Time:} In IoT environments, computation time and latency are very crucial. In this regard, the SDP \textit{processing time} is directly proportional to response time of the user requests and therefore we note that these metrics as native pipeline performance metrics. Processing time is measured in seconds, which is defined as the total time taken to process a data in a pipeline from source to destination. The source is a sensor node and sink is a storage/other end point in Cloud Infrastructure as described in Section~\ref{sec:probForm}. The processing time is addition of  both \textbf{communication} and \textbf{computation time} (latency). In the below paragraphs, we formulate the mathematical equations used to calculate these metrics. These metrics are calculated using logs of MinIO, MQTT, Apache NiFi and OpenFaaS gateway.  }
    
     \rtwo{
     \textbf{Computation time } : The computation time is calculated as summation of time required to compute a data unit by individual processing units (serverless functions) in a data pipeline. Let $R$ be the set of $n$ number of concurrent users requests $R = \{ r_1, r_2, \dots, r_n\}$. In our experiments the value of $n$ is considered from $10$ to $300$ and each user request carries a data unit to be processed by serverless function in the pipeline. The computation time of individual $i^{th}$ user request is 
    \begin{equation}\label{eq:comTime}
        P(r_i) = \sum_{\forall{f_j\epsilon F}} P(r_i,f_j)
    \end{equation}
    where $F=\{f_1,f_2,\dots,f_m \}$ is a set of $m$ number of functions and $P(r_i,f_j)$ represents the time taken by the function $f_j \in F$ to execute or process the request $r_i \in R$.
    }

    \rtwo{\textbf{Communication time}:  The  communication time denoted as $C_T(r_i)$ of $i_{th}$ user request is the time required to move data unit from source to sink  in pipeline excluding the computation time. It's summation of \textbf{disk access time} (for intermediate storage units) and \textbf{network communication time}. 
    }
    \rtwo{
     Let $D_{at}(r_i)$ be the arrival time at source and $D_{ct}(r_i)$  be the completion time of the $i^{th}$ user request at sink. The total duration of  serving the user request $D(r_i)$ is measured as
    \begin{equation}\label{eq:duTime}
    D(r_i) = D_{ct}(r_i) -  D_{at}(r_i)
    \end{equation}
    From equation~\eqref{eq:comTime} and equation~\eqref{eq:duTime}, the communication  time is calculated  as
    \begin{equation}\label{eq:coTime}
       C_T(r_i) = D(r_i) -  P(r_i)
    \end{equation}
      }
    \rtwo{Further, In the SDP approaches the data units are stored in the intermediate storage units and served to serverless functions for processing. The total time that data unit resides in the storage unit is considered as \textbf{disk access time} denoted as $DAT(r_i)$ is inclusive of  time required to store and access the data units. 
    Let $DU_{at}(r_i, S_j)$ and $DU_{dt}(r_i, S_j)$  be the arrival time  and departure time respectively of the $i^{th}$ user requests' data unit in the storage unit $S_j \in S$. The total duration of disk access time is calculated as 
    \begin{equation}
        \label{eq:dasTim}
        DAT(r_i) = \sum_{S_j \in S}{DU_{dt}(r_i,S_j) -  DU_{at}(r_i,S_j)}
    \end{equation}
    where $S = \{S_1,S_2,\dots,S_k\}$ is a set of $k$ number of storage units.
    }
    \rtwo{Finally, \textbf{network communication time} denoted as $NCT(r_i)$ is the summation of  time required  to move a data unit (of a user's request) in network from user's device to the sequence of processing units (serverless function) and intermediate storage units until the final data sink. Now from the equation~\eqref{eq:coTime} and equation~\eqref{eq:dasTim} network access time is calculated as 
      \begin{equation}
        \label{eq:nasTime}
        NCT(r_i) =  C_T(r_i) - DAT(r_i)
    \end{equation}
    }

\vspace{-5mm}
    \item \textbf{Average CPU utilization:} The average CPU utilization is measured in percentage (\%) and is calculated over time period from pipeline invocation till the data is received in the final destination. 
    \item \textbf{Average memory utilization:} It is measured in percent (\%) and is calculated as sum of total free memory, cache memory, memory in buffer and divided by total memory. Similarly, average disk utilization is measured in percentage (\%).
    \item \textbf{Network received:} This is calculated as bytes per second and is calculated as sum of bytes received on the network over a period of time. 
    \item \textbf{Network transmitted:} It is calculated as bytes per second and is calculated as sum of bytes uploaded on the network over a period of time.
    \item \textbf{Disk I/O Read:} This is measured in kilobytes and is calculated as sum of bytes read from file system over a period of time. 
    \item \textbf{Disk I/O Write:} It is measured as bytes per second and is calculated as sum of bytes written in to the file system over a period of time.
\end{itemize}
 


      



      

\subsection{Experimental Setup}
 
\begin{table*}[ht]
\centering
\caption{Hardware configuration for experimental setup}
\label{table:hwConfig}

\begin{tabular}{|l|l|l|l|}
\hline
\textbf{Device name} & \textbf{Configuration (Processor, RAM)} & \textbf{Quantity} & \textbf{Node type} \\ \hline
RPi 4B model      & Quad-CoreCortex A72, 4GB LPDDR4              & 2 & Fog node      \\ \hline
RPi 3B model      & Quad-CoreCortex A53, 4GB LPDDR4              & 1 & Gateway node  \\ \hline
Virtual machine   & 4-Core, 8GB DDR4                             & 1 & Cloud node    \\ \hline
Minix Neo Z64-W10 & Quad Core Z3735F (64 bit), 2GB DDR3          & 1 & Fog node      \\ \hline
Router            & Inteno DG200 model with 1000Mbps full duplex & 1 & Network layer \\ \hline
\end{tabular}
\end{table*}

The Docker Container Engine v19.03.12 is installed in both fog and cloud infrastructure in swarm mode. The OpenFaaS serverless platform is used as a FaaS engine configured in fog and cloud. OpenFaaS functions are developed using the programming language templates (bash streaming and Python 3.7). OpenFaaS command line interface (CLI) is used to build and deploy the functions into the OpenFaaS gateway. The Apache NiFi v1.3.2 is used in both fog and cloud as container service. The Apache NiFi user interface is used to design the data flow and monitor the flow files. The MinIO is deployed using docker compose service and volumes are mounted in the host machines. The MinIO client is used to create and configure the settings for event notifications on bucket.

A set of hardware devices and cloud resources are used to deploy and setup application services, as shown in Table~\ref{table:hwConfig}. Three Raspberry Pi 4B models and MiniX NEO Z83-4U Intel Mini PC are used for setting up fog infrastructure. For cloud infrastructure, the virtual machines of size \textit{m2.medium} with vCPU and 8GB RAM resembling similar capacity as AWS are provisioned from the University's private OpenStack cloud. The Raspberry Pi 3B model is used as a gateway node, and all the edge and fog devices are connected in a LAN with 1000 Mbps network bandwidth using Inteno DG200 router. The fog devices are connected to cloud services via 1000 Mbps network bandwidth. \rone{The network setup used for   interconnection between edge, fog and cloud environments are dedicated to these experiments}.  

Upon setting up of necessary hardware and application services, the use cases (Aeneas, PocketSphinx and custom video processing ) are deployed. The corresponding performance metrics are measured, and results are discussed in below subsections.


\subsection{Results and Discussion}
We considered scaling the number of users as a parameter to measure the performance of the approaches because the rate of concurrent arrival of user requests heavily impacts the pipeline performance. To measure the performance of all the metrics, several users are scaled from 1 to 15 for video applications (we used a chunk of video file as one user request) and 10 to 300 for Aeneas, PocketSphinx applications, and the corresponding SDP performances were measured. \rone{However, for calculating the processing time we considered 100 users in Aeneas and PocketSphinx due to data units were started dropping in MQTT based SDP.} 

\subsubsection{\textbf{Performance metrics observed with Aeneas application}}
\label{sec:results}
\begin{figure}
    \centering
    \includegraphics[width=0.95\linewidth]{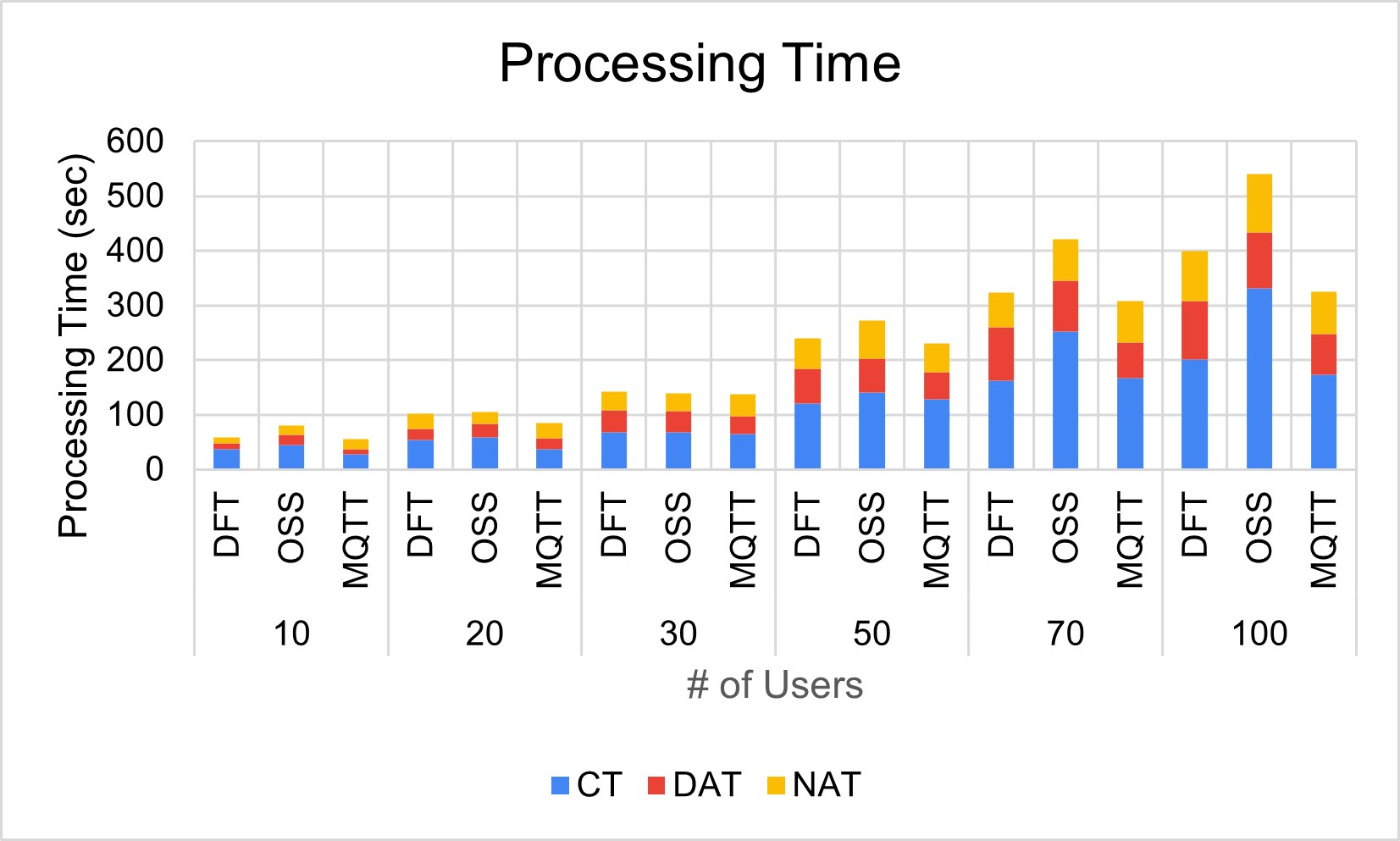}
    \caption{\textbf{\textit{\rtwo{Aeneas application - Processing time measured in seconds}}}}
    \label{fig:PT_Aenenas}
\end{figure}
The processing time was studied in all the three SDP approaches, as shown in the Figure~\ref{fig:PT_Aenenas}. Here, the y-axis  represents processing time in seconds, and the x-axis shows the \# of users.  The OSS requires a maximum of $540s$ to complete the $100$ users requests, whereas MQTT based SDP and DFT processed in $330s$ and $324s$, respectively.  The OSS had more processing time as the number of users increases, because MinIO notification invocations are synchronous. This pipeline had around five serverless functions as shown in Table~\ref{table:functions} and  two functions were extra to facilitate for retrieving the object data from MinIO buckets and this can lead to extra processing time. \rtwo{The computation time in OSS and MQTT based SDP was higher, where as disk access time was more in DFT. The internal queues in DFT manages efficient flow of data that makes to stay the data units in the queue that increase the DAS time. In OSS and MQTT based SDP, events were triggered as the data units arrived in to storage units which makes openfaas gateway to push these asynchronous user requests to NATs queue that increases the overall function execution time.
The average computation time was highest in OSS with $351$s but DFT had lesser computation with more disk access time of $280$s}. 


\begin{figure}
    \centering
    \includegraphics[width=0.95\linewidth]{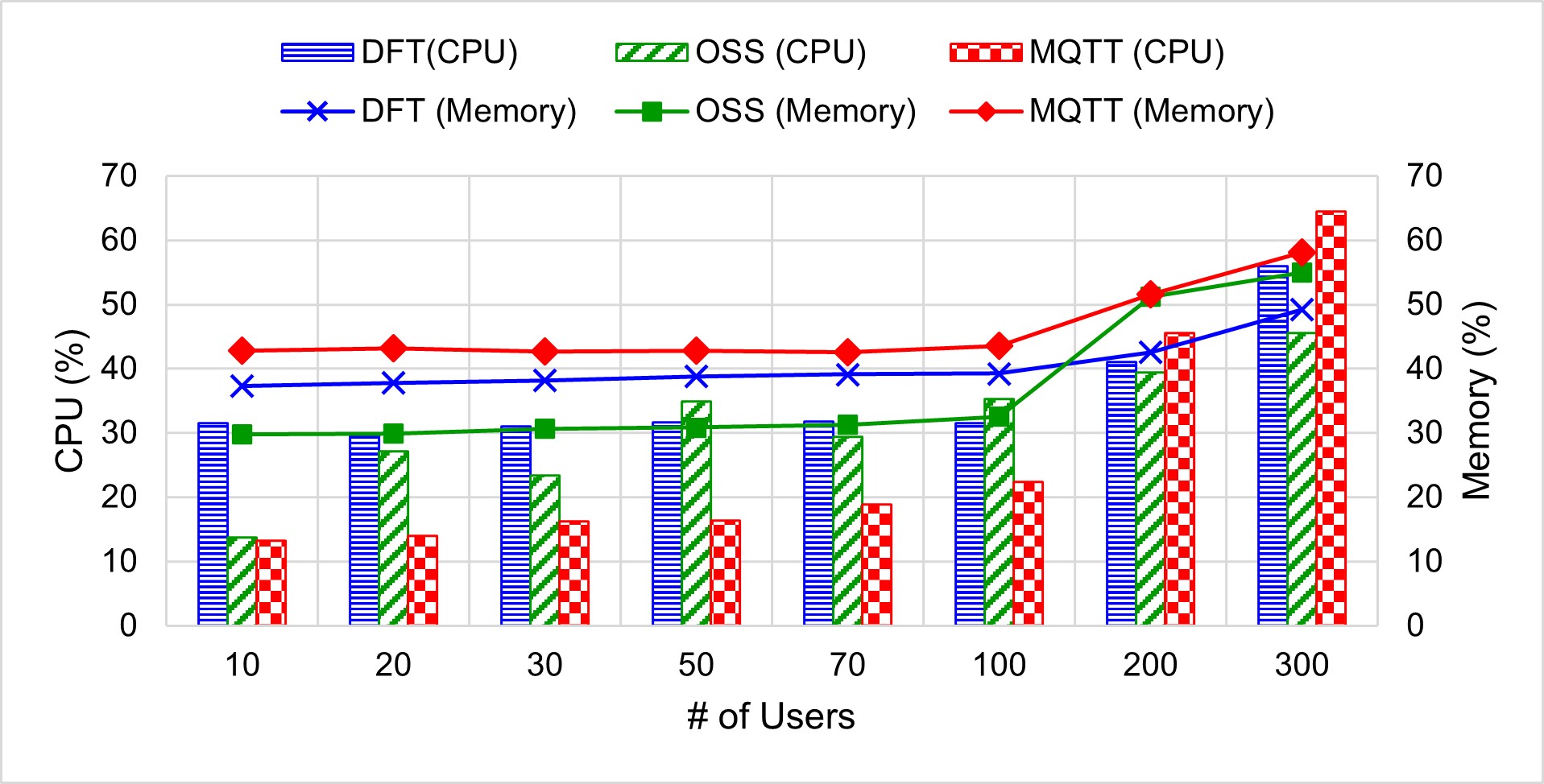}
    \caption{\textbf{\textit{\rone{Aeneas application }}} \rone{ - Average CPU and Memory utilization}}
    \label{fig:CPU_Memory_Aeneas}
\end{figure}

The average CPU utilization and Memory utilization were measured, as shown in the  Figure~\ref{fig:CPU_Memory_Aeneas}. The primary vertical y-axis shows an average CPU utilization measured in percentage (\%) and the secondary y-axis shows the Memory utilization. The DFT consumed highest CPU of $36\%$,  whereas MQTT based SDP consumed a lesser CPU of $21.06\%$ and OSS had moderate CPU utilization of $31\%$. But MQTT based SDP started using more CPU after 300 users request and further the data units in the pipeline started dropping. 

The Object Store and MQTT-based SDP used more number of lightweight python-based serverless functions, and DFT had a higher CPU utilization due to the set of Apache NiFi processors used in the pipeline that require extra computation power apart from serverless functions. 

The average Memory utilization at the secondary y-axis is measured in percent (\%). The MQTT-based SDP approach has the highest memory usage footprint of an average $45.92\%$, whereas DFT and MQTT based data pipelines used an average of $40.29\%$ and $36.42\%$ of memory, respectively. 
\begin{figure}
    \centering
    \includegraphics[width=0.95\linewidth]{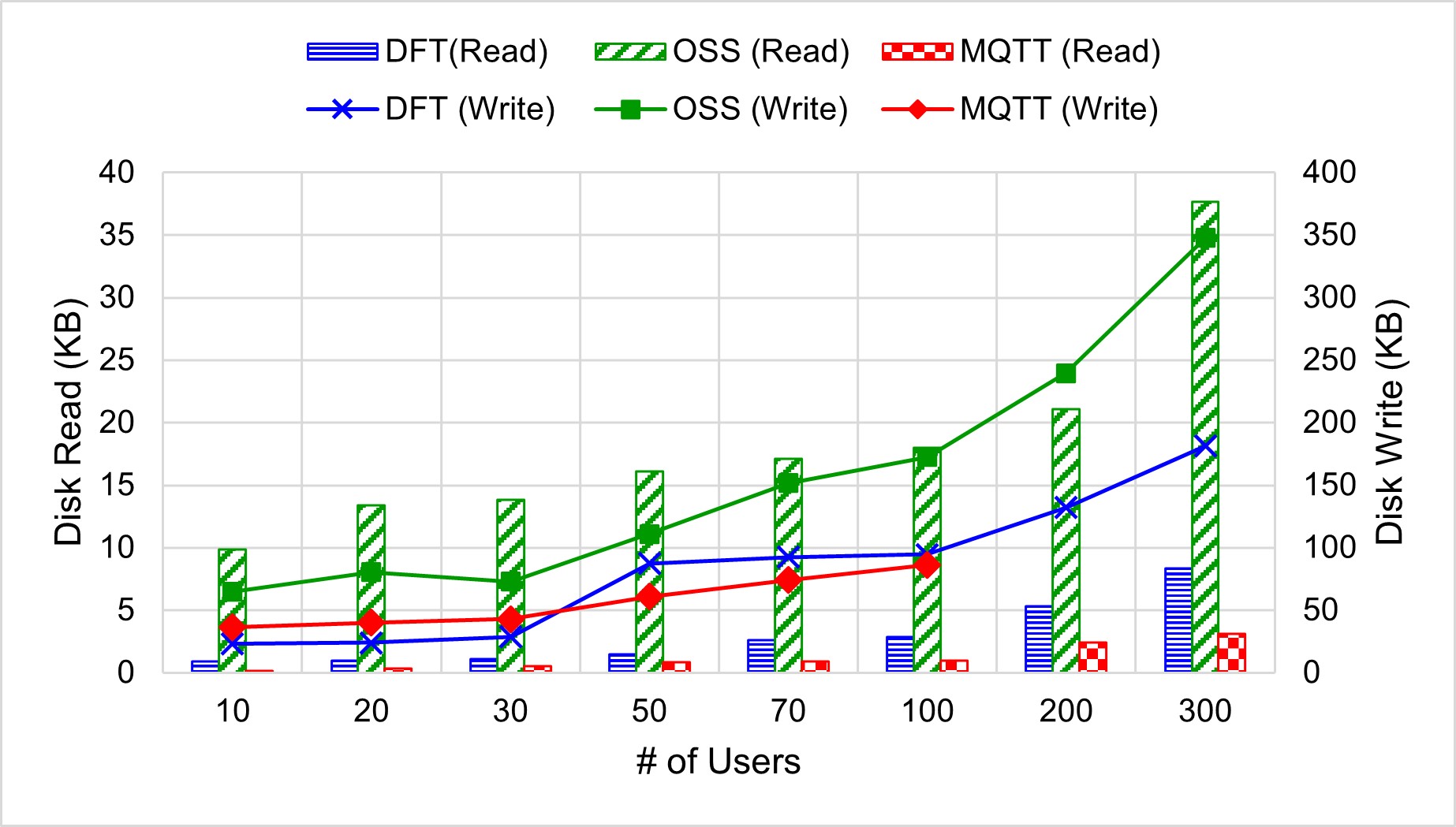}
    \caption{\textbf{\textit{\rone{Aeneas application }}} \rone{ - Average Disk Reads and average Disk Writes measured in Kilobytes}}
    \label{fig:Disk_Read_Write_Aeneas}
\end{figure}

In  the Figure~\ref{fig:Disk_Read_Write_Aeneas}, the primary y-axis represents disk I/O read and the secondary y-axis shows a disk I/O writes measured in Kilo Bytes (KB). In the case of the OSS SDP approach, $18.38$KB disk reads which is maximum as compared to DFT and MQTT with  $2.9$KB and $1.17$KB, respectively for 300 users.  

Similarly, OSS had a higher disk writes of $155$KB as compared with DFT and MQTT with $83$KB and $96$KB respectively for $300$ users. The OSS has more disk read/writes due to the read/write of bucket values based on each trigger. While in Apache NiFi, data flow is through the queue and doesn't had sever disk read/writes.

\begin{figure}
    \centering
    \includegraphics[width=0.95\linewidth]{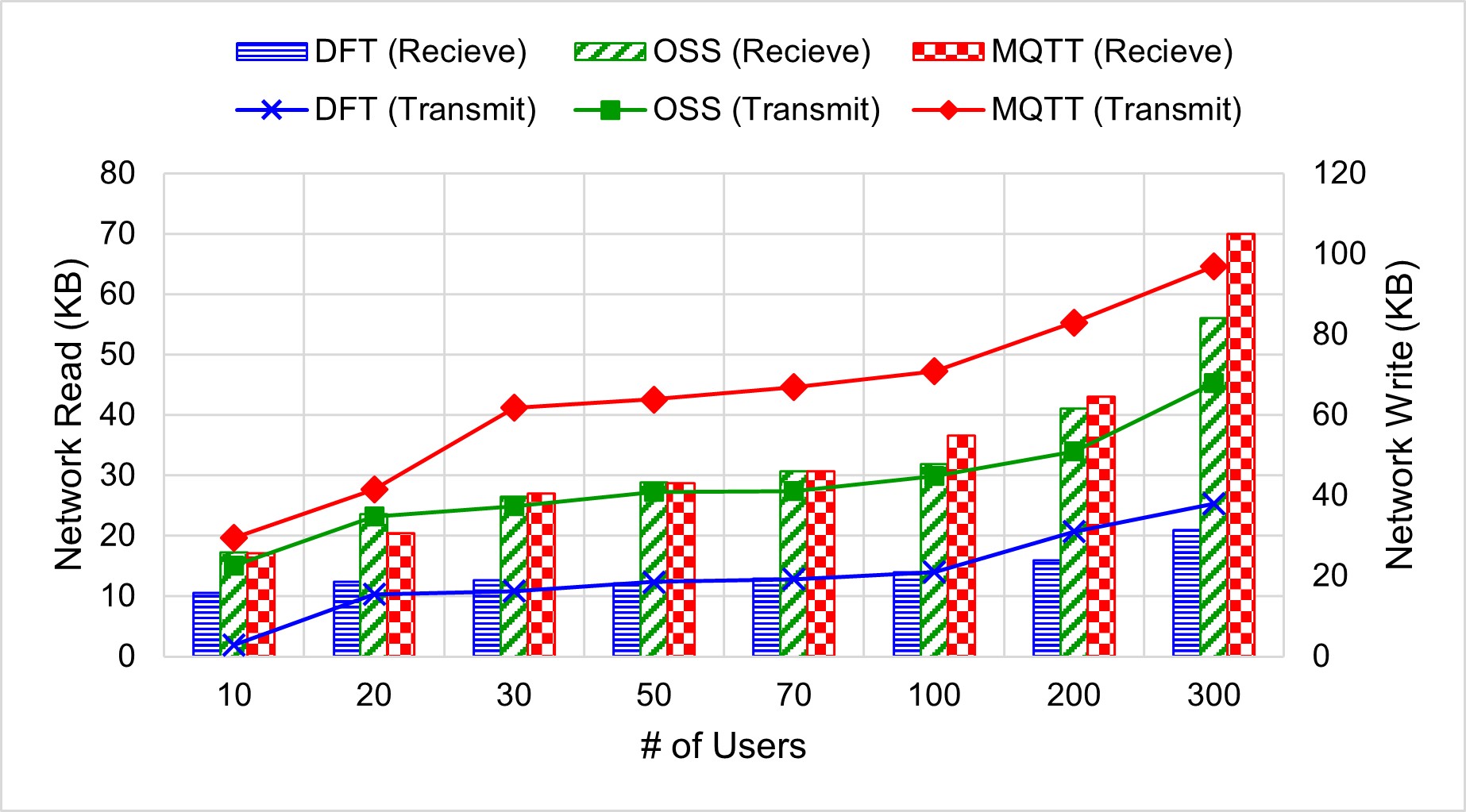}
    \caption{\textbf{\textit{\rone{Aeneas application}}} \rone{ - Average Network Transmit and Average Receive data and measured in Kilo Bytes}}
    \label{fig:NW_Read_Write_Aeneas}
\end{figure}
Network performances of the SDPs were measured as Network receive and transmit bytes as shown in Figure~\ref{fig:NW_Read_Write_Aeneas}. The OSS and MQTT has highest Network receive bytes calculated as average overall users $32$KB and DFT performed well with $34$KB. But for Network transmit bytes, MQTT had the highest reading with $64.3$KB for $300$ users. While, DFT had less network transmit bytes of $20$KB over $300$ users. MQTT recorded with  highest values in terms of  network performance due to each data flow with topic will be  published and subscribed over the network, even in OSS most of the network operations recorded on buckets with event triggers.     

\rone{In this application, MQTT based SDP had a lowest computation time with minimum  processing time as compared with DFT and OSS. Further, MQTT based SDP did not experience any drop of data units in the pipeline as compared with PocketSphinx and Custom video application. Moreover, CPU, Memory consumption and data Read/Writes metrics were also lowest, but there was raise in Network Receive/Transmit but it was negligible since the data unit size in the pipeline was very minimum. Considering the above metrics and associated SDP performances for Aeneas application, it is evident that  MQTT SDP worked better over OSS and DFT, as shown in suitability table Table~\ref{table:suitability}}. 

\subsubsection{\textbf{Performance metrics of the PocketSphinx application}}
\begin{figure}
    \centering
    \includegraphics[width=0.95\linewidth]{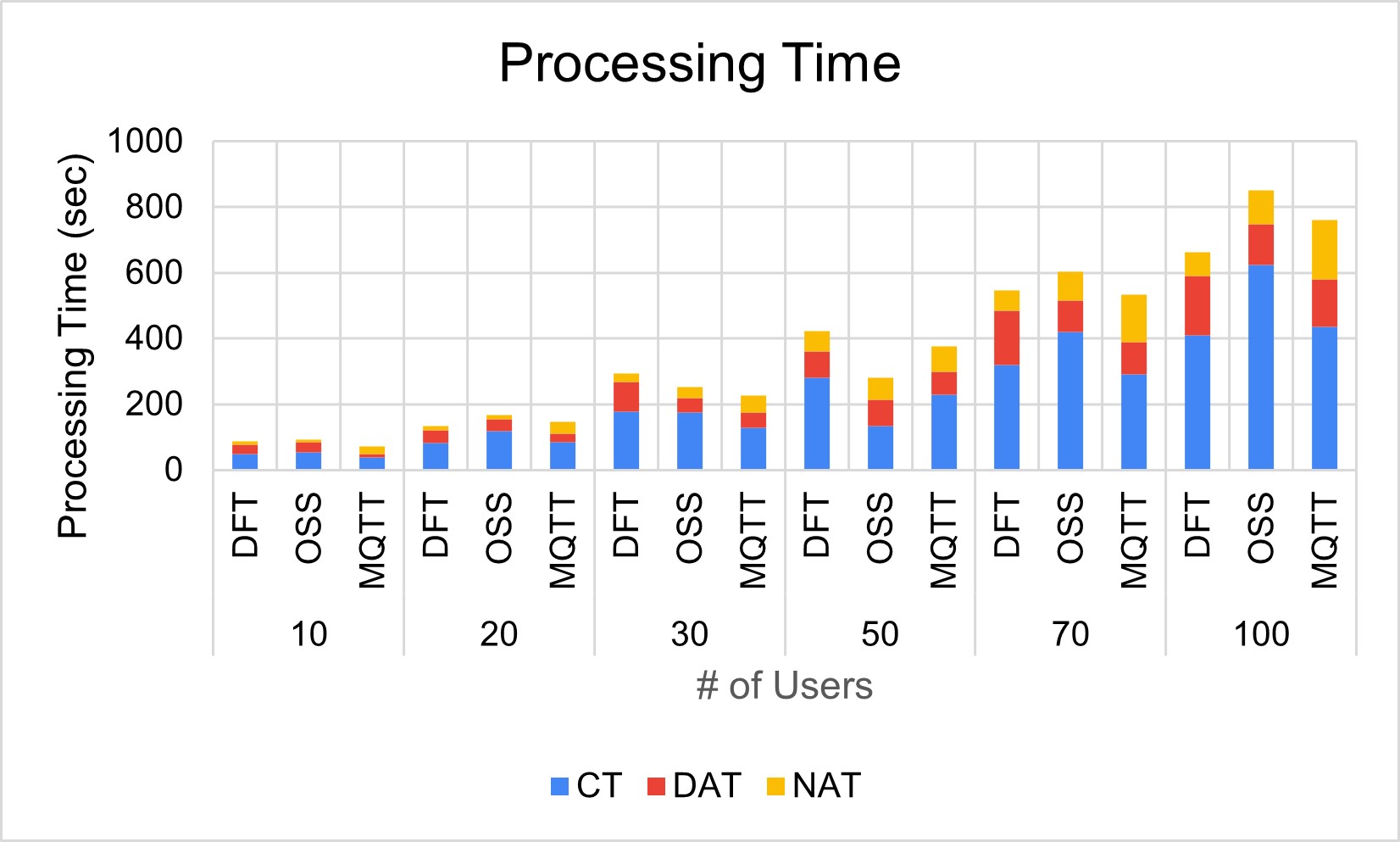}
   \caption{\textbf{\textit{\rtwo{PocketSphinx application}}} \rtwo{- Processing time measured in seconds}}
    \label{fig:PT_Pocketsphinx}
\end{figure}

The processing time of the PocketSphix application over all the proposed SDP's were measured as shown in  Figure~\ref{fig:PT_Pocketsphinx}. The OSS had the highest processing time of $851s$, whereas DFT had $663s$ with minimum processing time. This application had large a set of functions in the pipeline and OSS event notifications are set to three buckets to store intermediate results. All the events are triggered asynchronously and lead to a larger processing time. \rtwo{As similar to Aeneas, DFT  shared highest disk access time, whereas OSS and MQTT had maximum computation time. In MQTT, major challenge was the data unit drop rate increased as the number of user requests increased and here, drop rate was  approximately $2\%$.}

\begin{figure}
    \centering
    \includegraphics[width=0.95\linewidth]{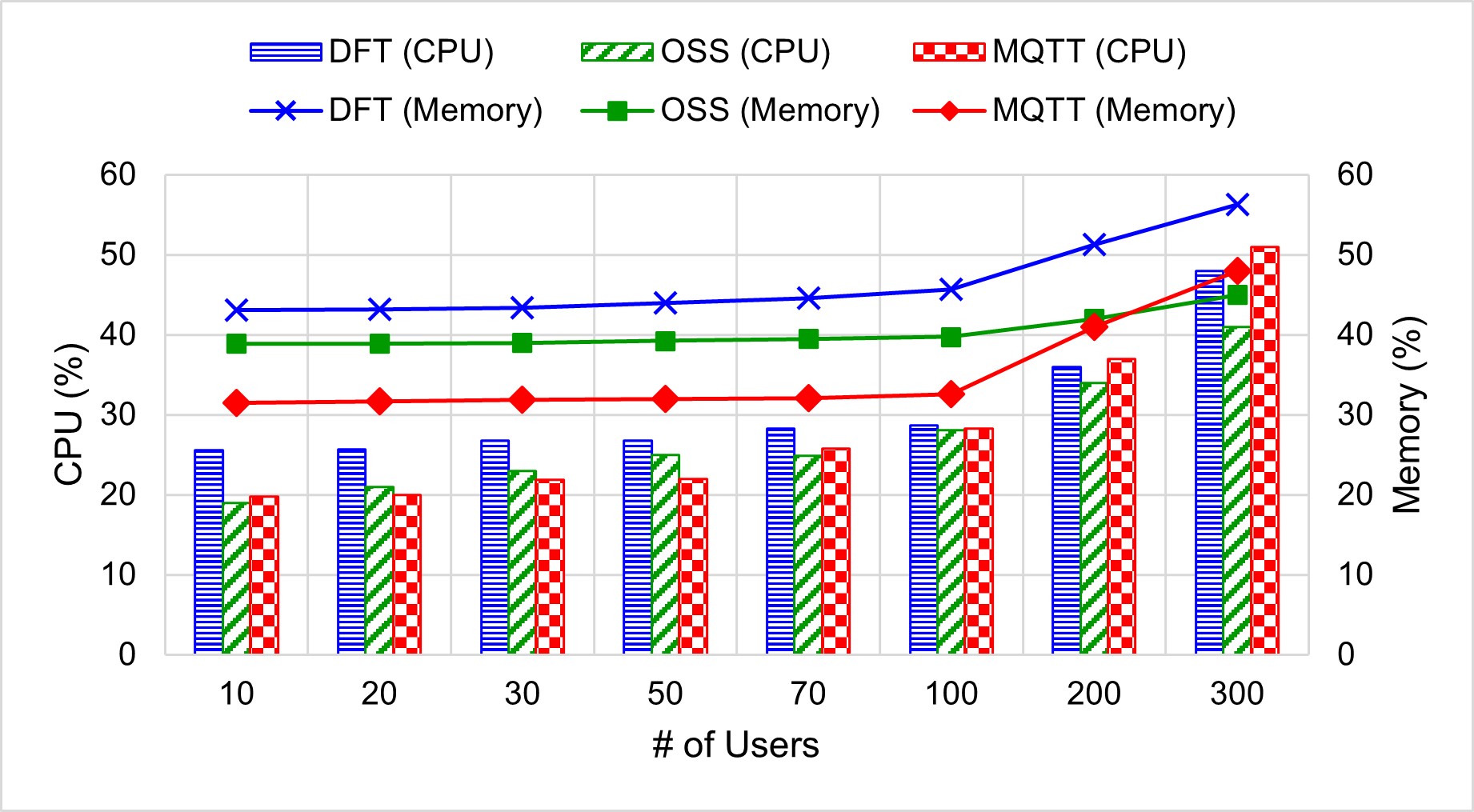}
    \caption{\textbf{\textit{\rone{PocketSphinx application }}} \rone{ - Average CPU utilization and average Memory utilization  and measured in \%}}
    \label{fig:CPU_Memory_pocketsphinx}
\end{figure}
The CPU utilization and Memory utilization were observed in \% as shown in Figure~\ref{fig:CPU_Memory_pocketsphinx}. The OSS and MQTT equally consumed the CPU of $28\%$ calculated over $300$ users as shown in the primary y-axis, whereas DFT consumed highest CPU $31\%$. However, MQTT had more CPU usage when user requests increased which is not suitable in terms of compute intensive and heavy compute bounded workloads. Similarly as in Aeneas, DFT uses Apache NiFi and required more CPU to execute processors concurrently. The memory utilization shown in the secondary y-axis,  DFT had highest average memory utilization of $46\%$ whereas MQTT used the highest  memory of $48\%$ after $300$ users.

\begin{figure}
    \centering
    \includegraphics[width=0.95\linewidth]{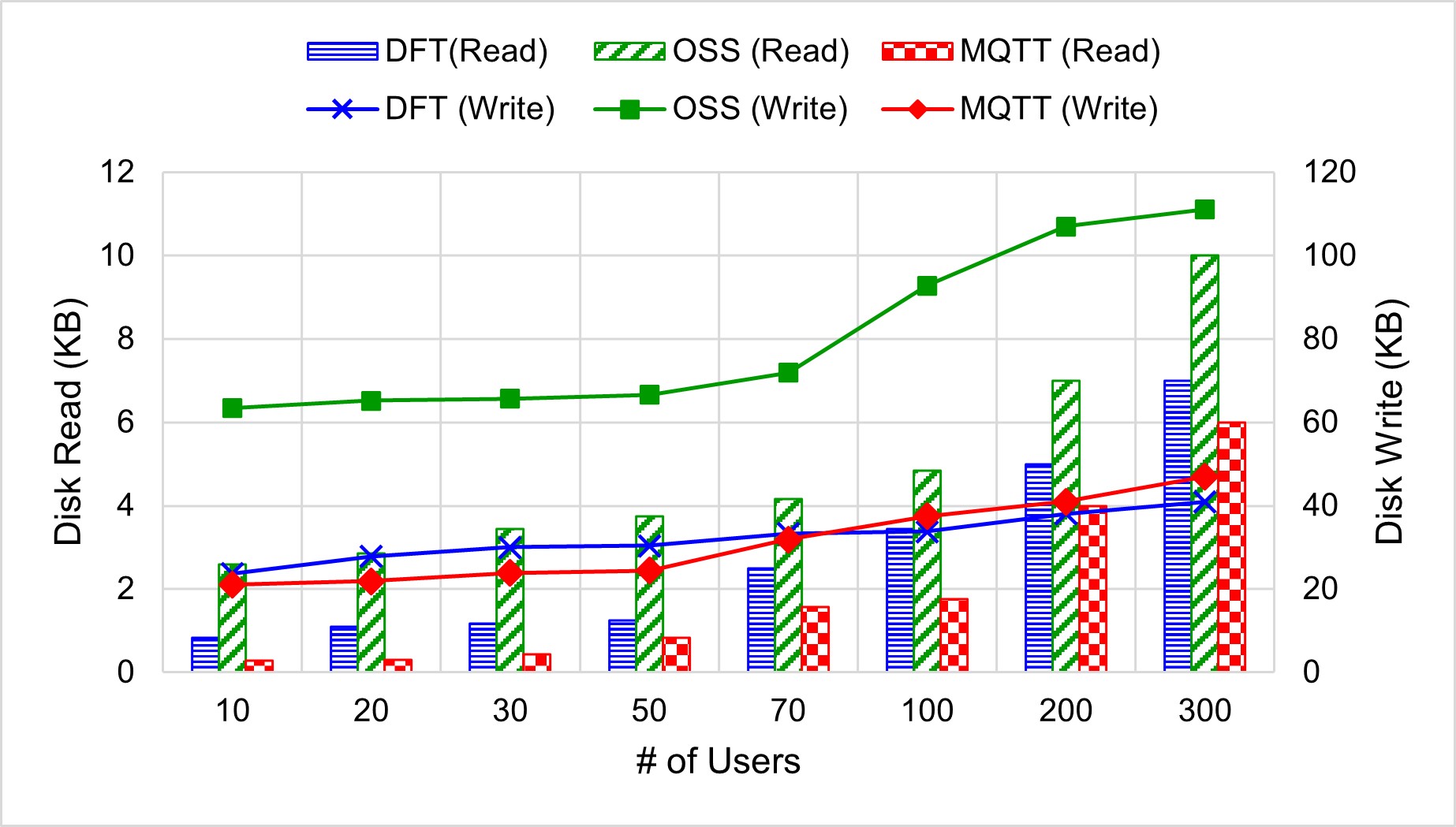}
    \caption{\textbf{\textit{\rone{PocketSphinx application} }}\rone{  - Average Disk Reads and average Disk Writes and measured in Kilo Bytes}}
    \label{fig:Disk_Read_Write_pocketsphinx}
\end{figure}
The disk I/O read and writes are measured in KB as shown in Figure~\ref{fig:Disk_Read_Write_pocketsphinx}. The OSS leads to more disk read $5$KB as compared to DFT with $2$KB and least with MQTT-based SDP of $1$KB over $300$ users. Similarly as in Aeneas, OSS utilizes the MinIO (S3) storage and at each event triggers on bucket notification, data would read from the disk leading to higher disk reads. The disk writes were shown in secondary y-axis, as similar OSS had significant disk writes of $70$KB because the number of buckets used were more as compared with Aeneas and Video processing application. The least disk writes by MQTT-based SDP and moderately by DFT is due to data pushed and pulled in the queue neither directly written nor read from the disk.
\begin{figure}
    \centering
    \includegraphics[width=0.80\linewidth]{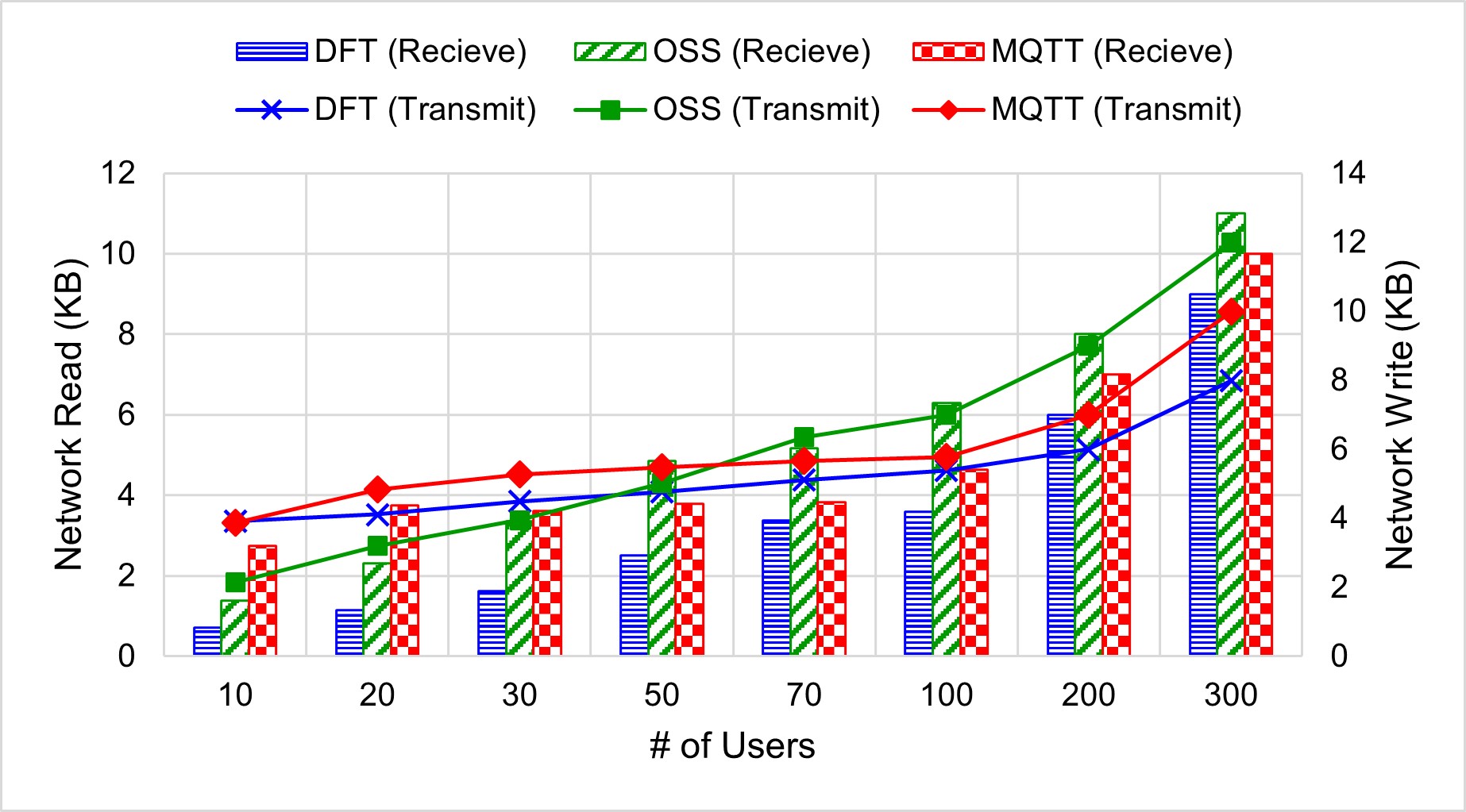}
    \caption{\textbf{\textit{\rone{PocketSphinx application }}} \rone{ - Average Network receive and transmit and measured in Kilo Bytes}}
    \label{fig:NW_Read_Write_pocketsphinx}
\end{figure}

The network receives and transmits were measured in KB as shown in Figure~\ref{fig:NW_Read_Write_pocketsphinx}. Here, the network received bytes performance of OSS had highest as $6$KB while DFT had least with $2$KB. The secondary y-axis represents the Network transmit bytes, similarly as above OSS had the highest transmit bytes with $7$KB and DFT had least with $5$KB over $300$ users. Push events and notifications events of MinIO make OSS to consume more Network receive and transmit bytes. Evermore, Pocketsphinx uses more buckets as compared with other applications.

Considering the above performance metrics, its observed that DFT and MQTT performed equally better on Pocketsphinx application as compared with OSS. However, Pocketsphinx required more bandwidth to transfer audio files over the fog network and this motivates to consider DFT as suitable SDP shown in Table~\ref{table:suitability}.
\subsubsection{\textbf{Performance metrics observed for custom video application}}
\begin{figure}
    \centering
    \includegraphics[width=0.95\linewidth]{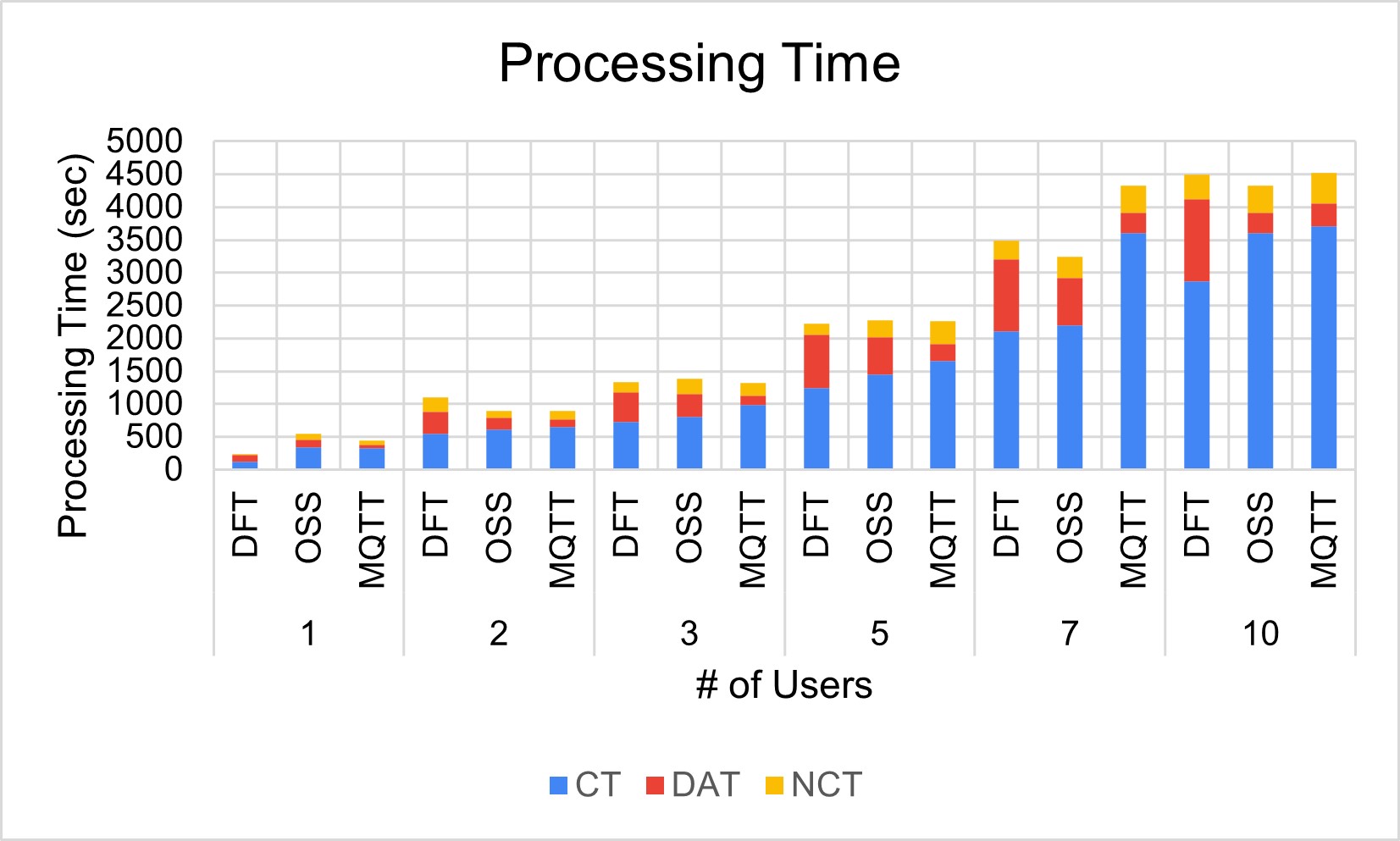}
    \caption{\textbf{\textit{\rtwo{Video processing application}}}\rtwo{ - Processing time measured in seconds}}
    \label{fig:PT_yolo_users_fps}
\end{figure}
The video processing application naturally demands huge computation power and more bandwidth to process and offload the video files. The quality of the video is determined by frame rate. A frame per second (fps) is the speed at which individual still images, known as frames, are displayed in the video. The higher value of fps in the video requires more resources to process. So, it is significantly necessary to investigate the performance metrics based on the change of the fps values. In our work, we considered fps values scaling from $1$ to $15$ and measured the performance. Along with this, its essential to investigate the rate of arrival of such user videos as measured in earlier applications. So, in this section, we will describe the performance metrics collected based on the change in fps values and arrival rate of user videos.

The processing time was measured in seconds (s) as shown in Figure~\ref{fig:PT_yolo_users_fps}. Here, the primary x-axis shows the number of users  The primary y-axis represent the Processing time measured for number of users. The DFT worked better with $4500s$ as compared with OSS and MQTT based SDP with $5520s$, $4515s$ respectively over $10$ users. \rtwo{As like other applications MQTT had major issue of dropping the data units intermediate pipeline and it was approximately $28\%$. The other challenge was, MQTT openfaas-connector invokes the function carrying heavy data input, which makes maximum openfaas NATs queue memory utilization and this raises an exceptions from openfaas gateway. Similar to other applications, computation time was maximum in OSS where as DFT has minimum disk access time.  The event triggers in MinIO and topic publish and subscription with huge multi-media (audio) data took quite higher processing time}.
\begin{figure}
    \centering
    \includegraphics[width=0.95\linewidth]{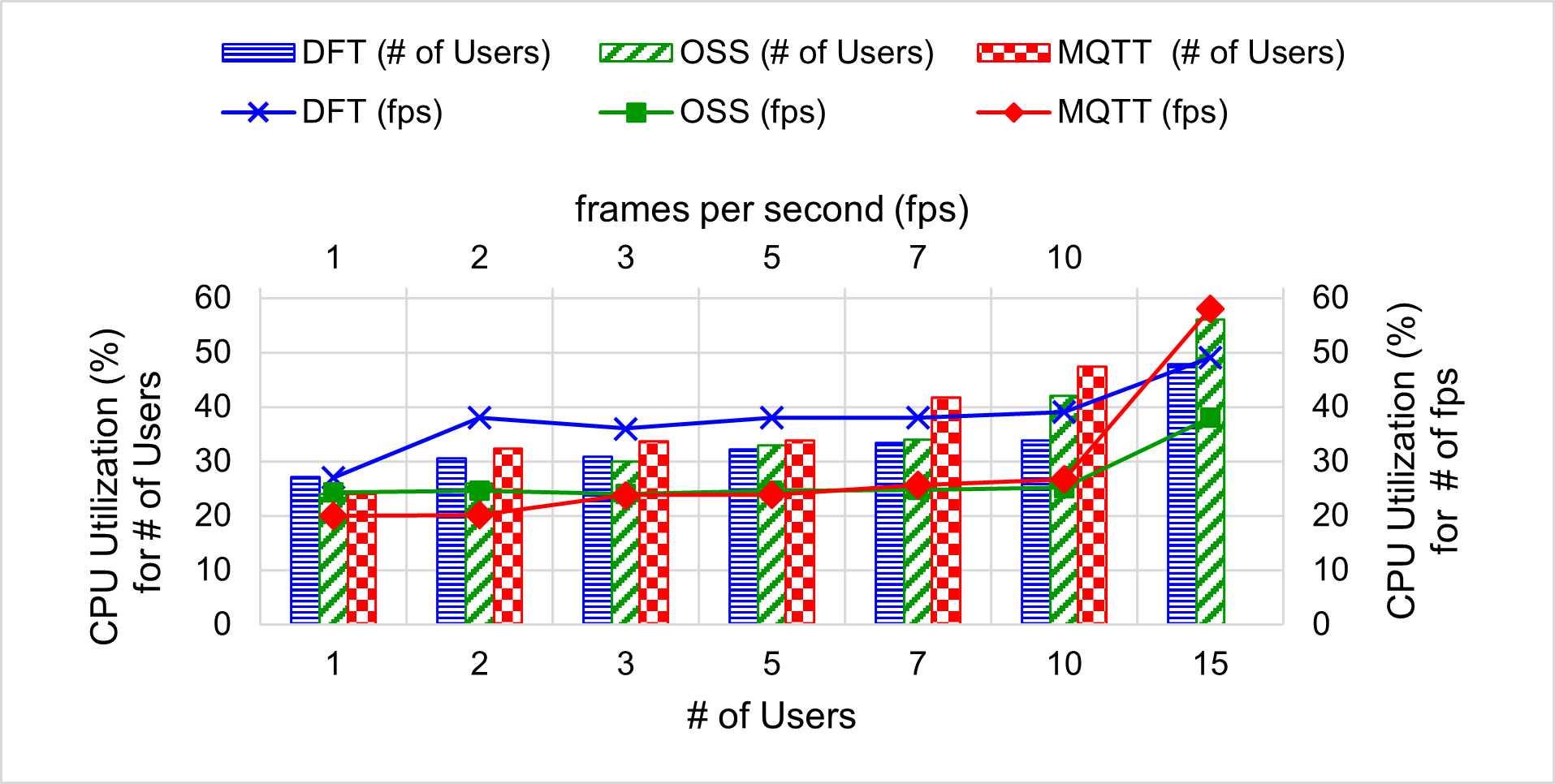}
    \caption{\textbf{\textit{\rone{Video processing application}}} \rone{- CPU utilization measured in percent (\%)}}
    \label{fig:CPU_yolo_users_fps}
\end{figure}


The Figure~\ref{fig:CPU_yolo_users_fps} represents the CPU utilization across scaling of number of users and fps values.
Based on scale of users, DFT performed better with CPU utilization of $32\%$, but MQTT consumed more CPU with $35\%$ over $15$ users. However, MQTT worked better based on the scale of fps values with $23\%$ and DFT consumed more CPU with $36\%$.
Considering both of the scenarios, OSS worked well. Even though the MQTT-based SDP works well in case of fps based scenario but consumed more CPU in another scenario.  

\begin{figure}
    \centering
    \includegraphics[width=0.95\linewidth]{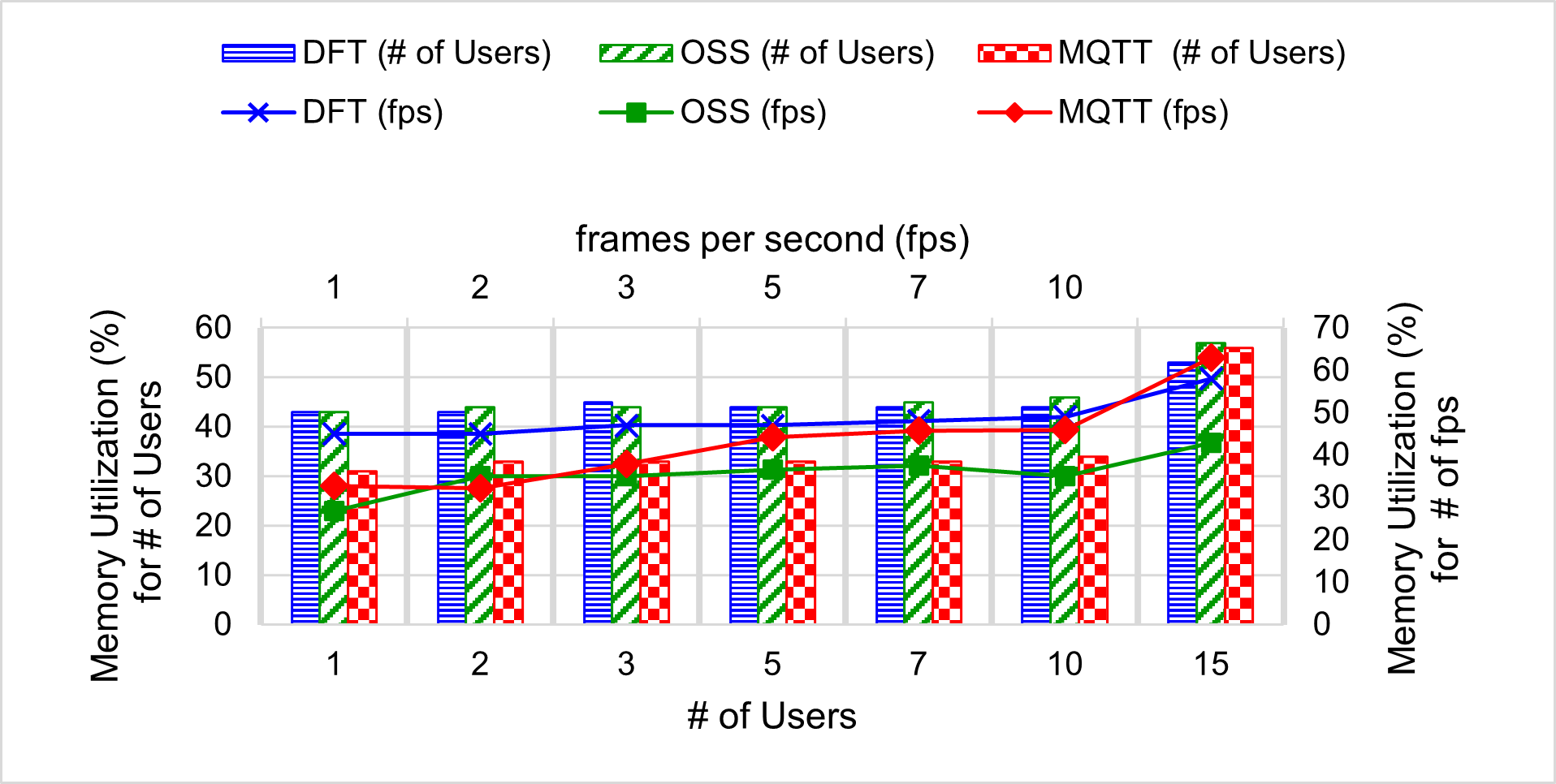}
   \caption{\textbf{\textit{\rone{Video processing application}}} \rone{- Memory utilization measured in percent (\%)}}
    \label{fig:Memory_yolo_users_fps}
\end{figure}

The memory utilization was shown in Figure~\ref{fig:Memory_yolo_users_fps}, MQTT based SDP consumed less memory as compared to OSS and DFT with $34\%$, $46\%$ and $44\%$ respectively over $15$ users. Interestingly, OSS consumed less memory with $35\%$ as compared with MQTT and DFT with $46\%$ and $49\%$ respectively. The MQTT-based SDP and OSS were good in terms of memory consumption considering both of the scenarios.
\begin{figure}
    \centering
    \includegraphics[width=0.95\linewidth]{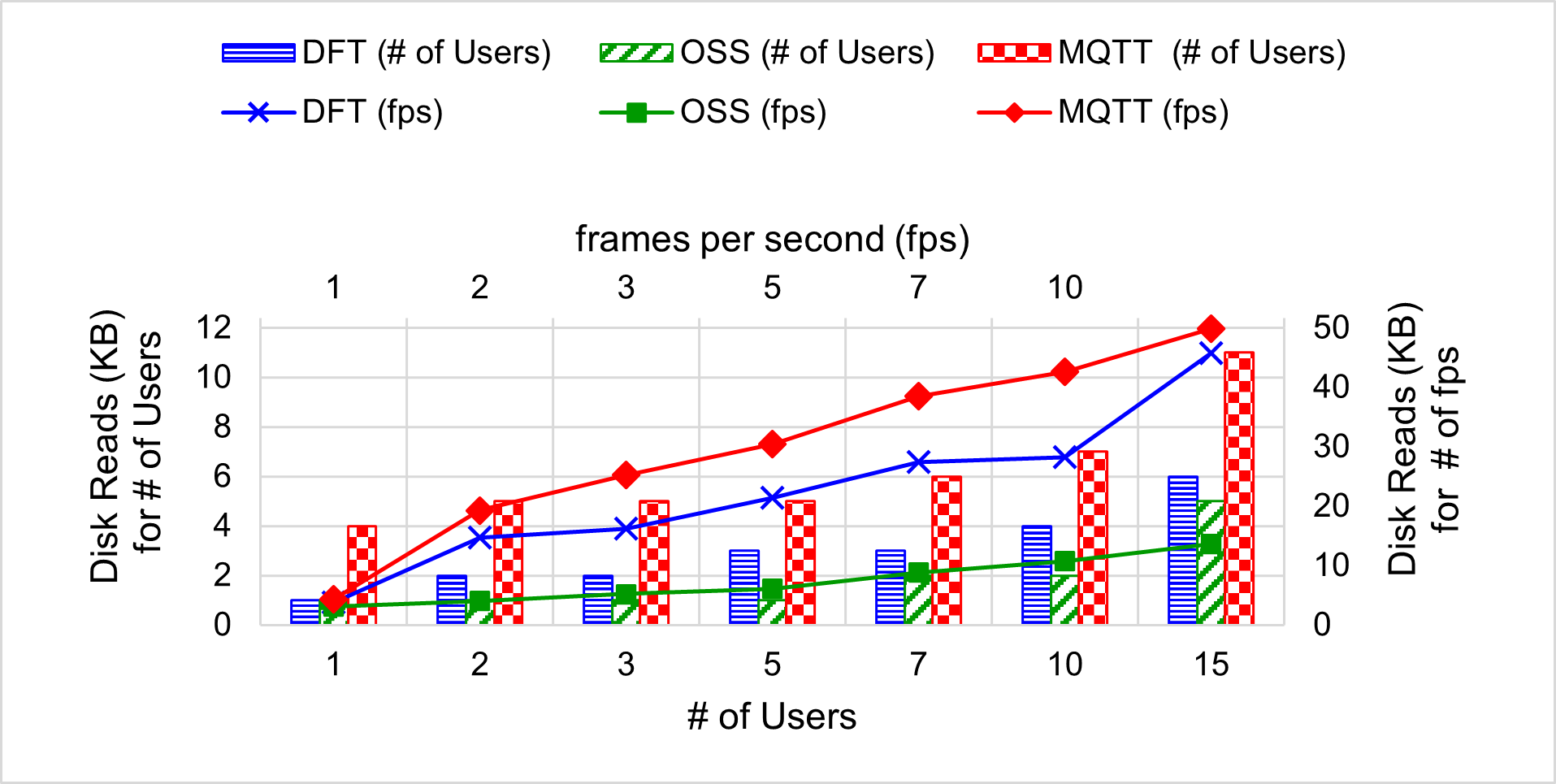}
    \caption{\textbf{\textit{\rone{Video processing application}}} \rone{- Disk Reads measured in Kilo Bytes (KB)}}
    \label{fig:Disk_Reads_yolo_users_fps}
\end{figure}

Disk Read for both of the scenarios shown in Figure~\ref{fig:Disk_Reads_yolo_users_fps}, OSS had very few disk reads in both of the scenarios with average values of $1$KB, $6$KB respectively. MQTT-based SDP had more disk reads and DFT moderately worked better in both of the scenarios. 
\begin{figure}
    \centering
    \includegraphics[width=0.95\linewidth]{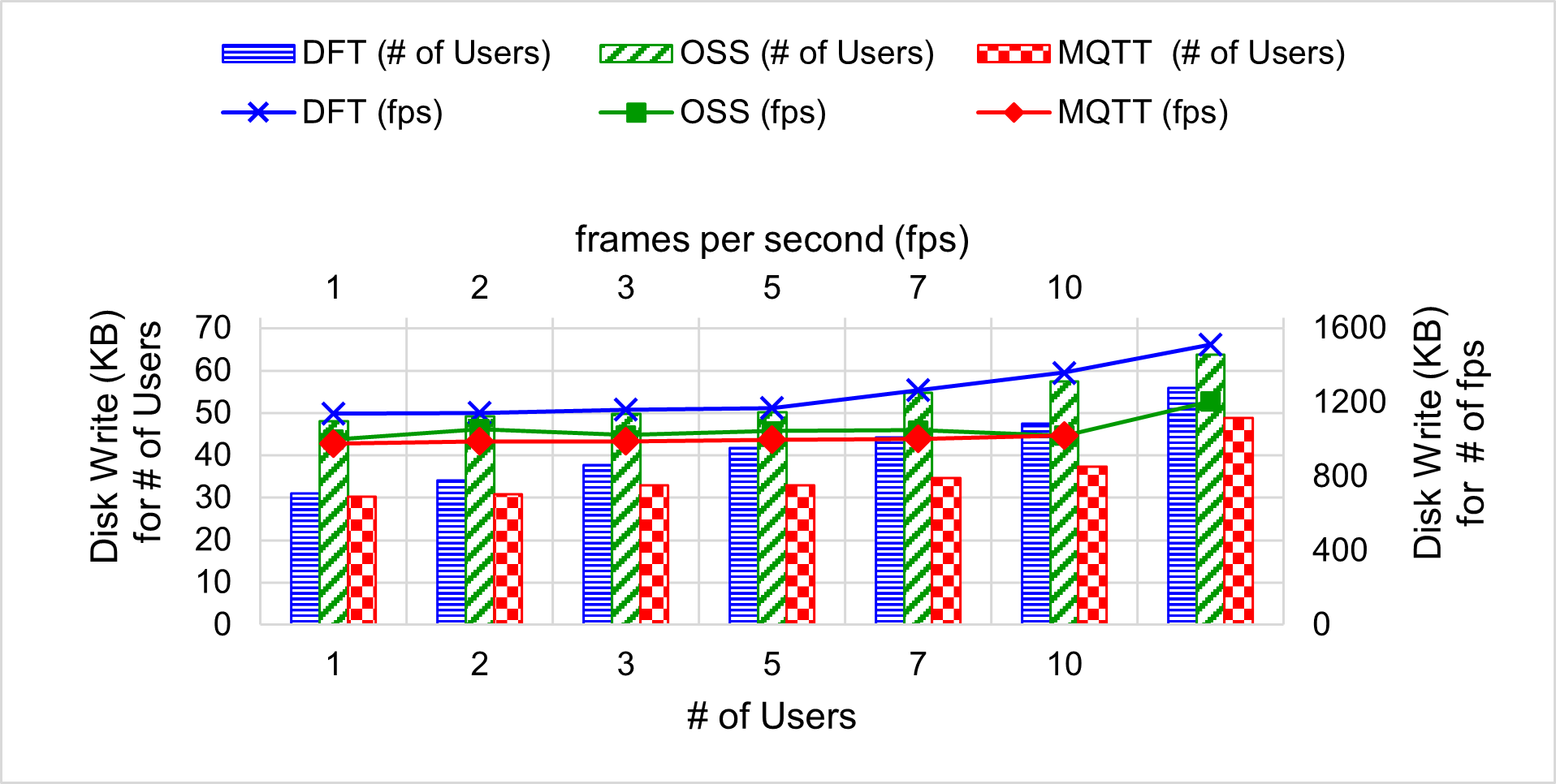}
    \caption{\textbf{\textit{\rone{Video processing application}}} \rone{- Disk Writes measured in Kilo Bytes (KB)}}
    \label{fig:Disk_Write_yolo_users_fps}
\end{figure}

Figure~\ref{fig:Disk_Write_yolo_users_fps} shows the Disk Writes, MQTT has minimum disk write in both scenarios with an average value of  $966$KB, $33$KB respectively. OSS had the highest disk writes based on the number of users while DFT had more based on the fps. The OSS will have obviously higher disk writes due to objects stored on the disks and DFT had more disk writes due to interaction of NiFi processors with file system.
\begin{figure}
    \centering
    \includegraphics[width=0.95\linewidth]{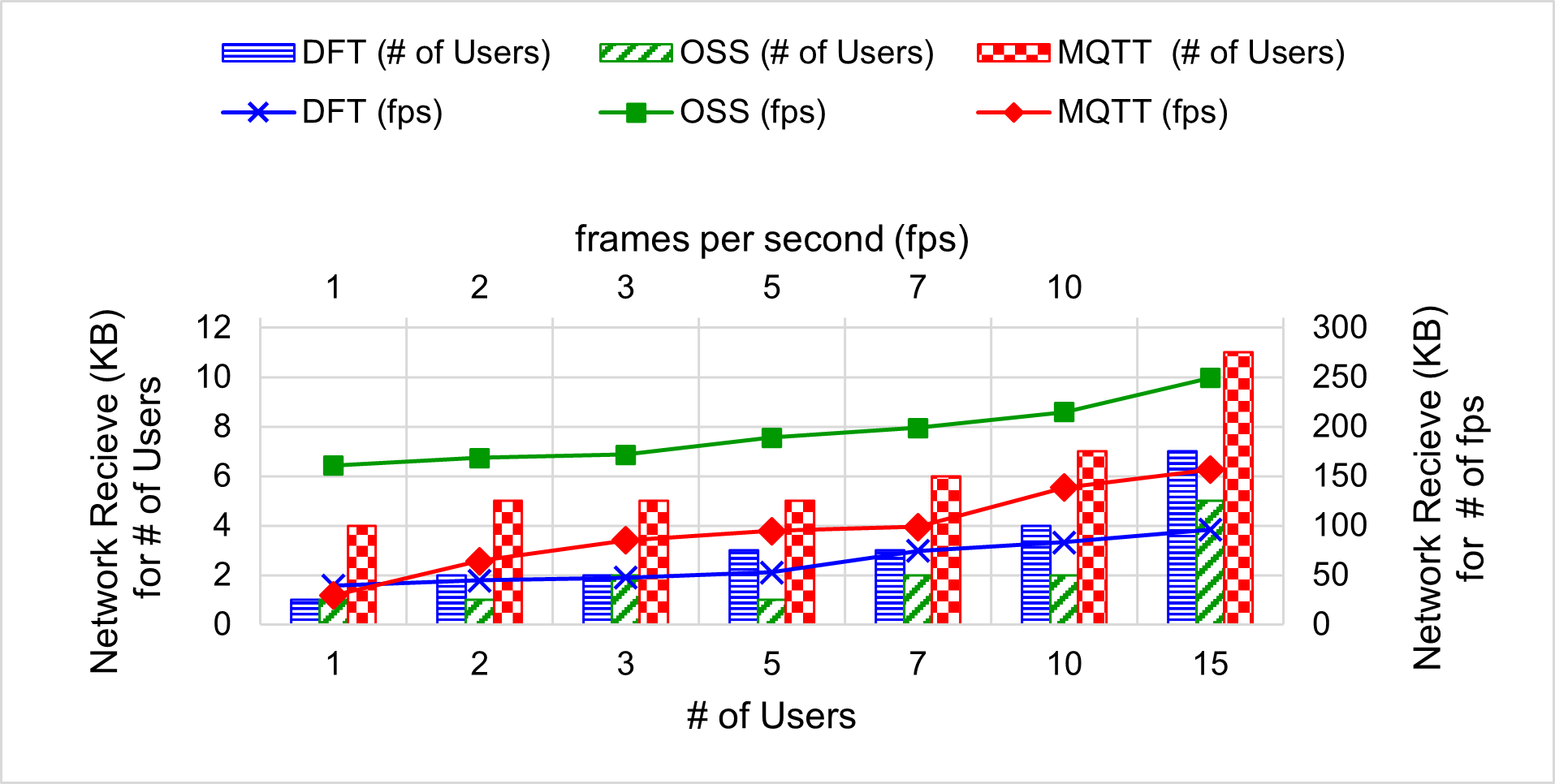}
    \caption{\textbf{\textit{\rone{Video processing application}}} \rone{- Network receive bytes measured in Kilo Bytes (KB)}}
    \label{fig:NW_Recieve_yolo_users_fps}
\end{figure}
\begin{figure}
    \centering
    \includegraphics[width=0.95\linewidth]{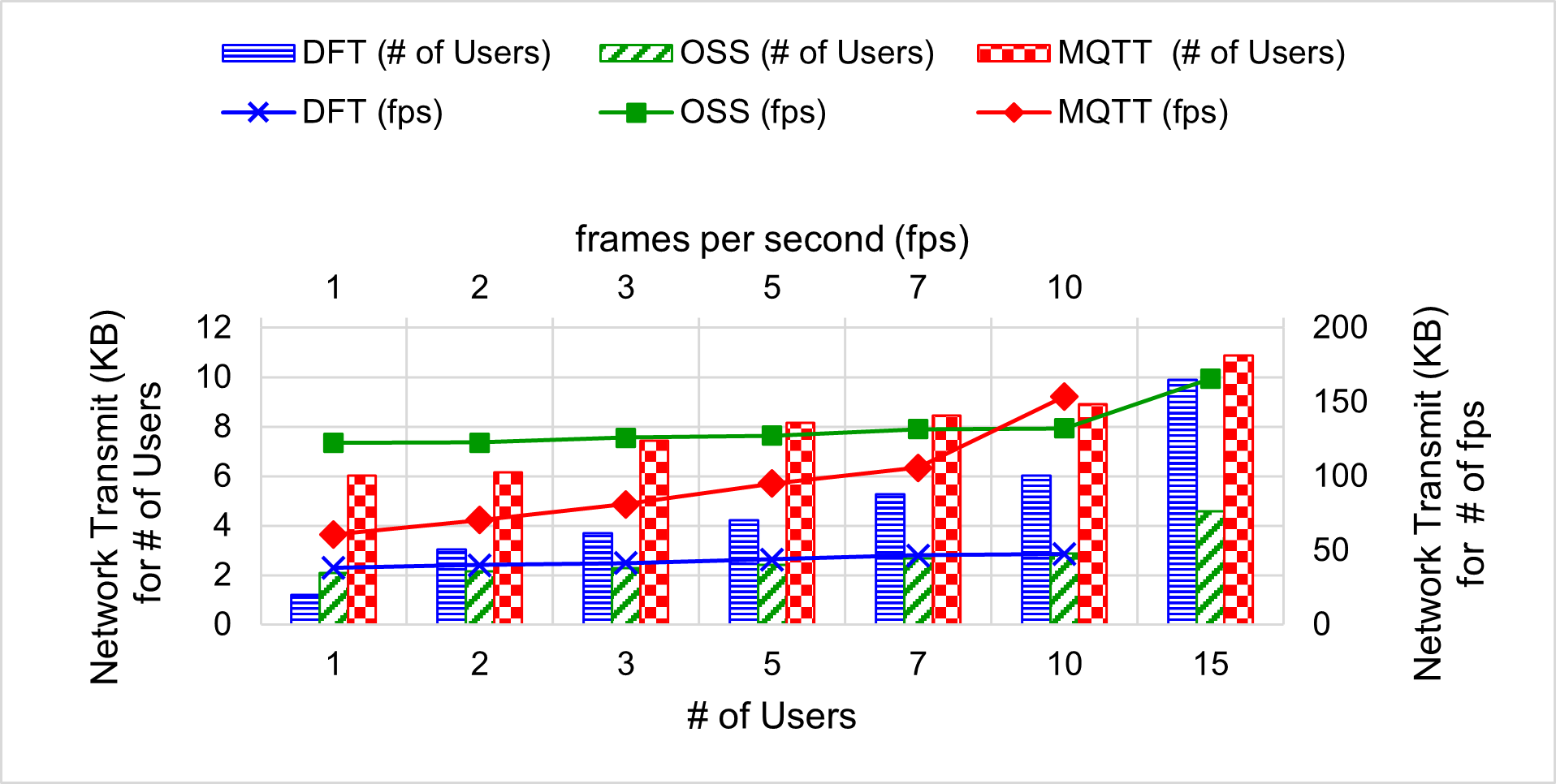}
    \caption{\textbf{\textit{\rone{Video processing application}}} \rone{- Network Transmit bytes measured in Kilo Bytes (KB)}}
    \label{fig:NW_Transmit_yolo_users_fps}
\end{figure}

The Figure~\ref{fig:NW_Recieve_yolo_users_fps} and Figure~\ref{fig:NW_Transmit_yolo_users_fps} represents the network performance  measurement for both of the scenarios in Kilo Bytes (KB). In terms of network receive bytes in scaling of users scenario, OSS worked very well with $2$KB and DFT moderately better with $4$KB, but MQTT had more network receive bytes with $5.3$KB considering $15$ users. However, the scaling of fps values scenario, DFT worked well with $59$KB, but OSS had maximum values over all the SDPs with $215$KB. 
In network transmit bytes, DFT performance was moderately good with $6$KB, $47$KB but OSS consumed minimum network resources(receive bytes and transmit bytes) in terms of  scaling the users whereas MQTT consumed more network resources. The MQTT based SDP consumed more network bandwidth due to publish of the multi-media data over the tcp network and mqtt-openfaas client always listen to this topics leading to higher network consumption. 

Considering the various performance metrics based on scaling of users and fps values, its observed that Video processing application consumes more CPU due to ffmpeg and YOLOv3 tools and even demand for more bandwidth to offload the multi media files between edge/fog/cloud infrastructure. However, DFT consumes less network resources but more CPU and disk resources, rather  OSS uses less CPU but required more network resources. MQTT based SDP  performance shows that its not suitable for Video processing due to heavy usage of resources. 

\subsubsection{\textbf{Suitability analysis }}

\begin{table*}[ht]
\centering

\caption{Average performance metric values and suitability index calculated across each application}

\resizebox{\textwidth}{!}{
\label{table:suitability}
\begin{tabular}{|l|l|l|l|l|l|l|l|l|}
\hline
Metrics/Application &
  \multicolumn{2}{l|}{Aeneas } &
  \multicolumn{2}{l|}{Pocketsphinx } &
  \multicolumn{4}{l|}{Video Processing application} \\
                          & \multicolumn{2}{l|}{}       & \multicolumn{2}{l|}{}   & \multicolumn{4}{l|}{}                                         \\ \cline{6-9} 
                          & \multicolumn{2}{l|}{}       & \multicolumn{2}{l|}{}   & \multicolumn{2}{l|}{\#   of Users} & \multicolumn{2}{l|}{fps} \\ \hline
Average Metric   Values\footnote{Average across all the applications with corresponding SDP }   & Min of Avg.                  & Max of Avg. & Min of Avg.        & Max of Avg.      & Min of Avg.            & Max  of Avg.           & Min  of Avg.       & Max   of Avg.     \\ \hline
Processing   Time (s)    & MQTT (MQ)             & OSS  & DFT          & MQ        & DFT               & OSS              & OSS          & MQ         \\ \hline
CPU Utlization   (\%)     & MQTT (MQ)             & DFT  & OSS/MQ       & DFT        & OSS               & MQ              & OSS          & DFT         \\ \hline
Memory   Utilization (\%) & OSS & MQ  & MQ          & DFT        & MQ               & OSS              & OSS          & DFT         \\ \hline
Disk Read (KB)            & MQTT (MQ)             & DFT  & MQ          & OSS        & OSS               & MQ              & OSS          & MQ         \\ \hline
Disk Writes   (KB)        & OSS  & DFT  & MQ          & OSS        & MQ               & OSS              & OSS          & MQ         \\ \hline
Network   Receive (KB)    & DFT       & OSS  & DFT          & MQ        & OSS               & MQ              & DFT          & OSS         \\ \hline
Network   Transmit (KB)   & DFT        & MQ  & OSS          & MQ        & OSS               & MQ              & DFT          & OSS         \\ \hline
Suitability index   (\%) &
  MQ (43\%) &
  DFT (43\%) OSS (43\%) &
  MQTT (57 \%) &
  MQTT (43 \%) &
  OSS (57\%) &
  MQTT (57\%) &
  OSS (71\%) &
  MQTT (43\%) \\ \hline
Suitable SDP              & \multicolumn{2}{l|}{MQTT}     & \multicolumn{2}{l|}{DFT} & \multicolumn{2}{l|}{OSS}            & \multicolumn{2}{l|}{OSS}  \\ \hline
\end{tabular}
}
\end{table*}
All the proposed SDPs were implemented for three fog computing applications, the observed performance of all metrics vs all applications are reported in Table~\ref{table:suitability}.
The focus of suitability analysis was to extract insights from an observed experimental results  in selection of best fit SDP for Aeneas, PocketSphinx and Video processing application. The overall results were summarized with  suitability index and  corresponding suitable SDP for each application was presented in the  Table~\ref{table:suitability}.

In this Table~\ref{table:suitability}, average performance metric values were calculated by averaging the recorded values of SDPs performance across individual application and then minimum and maximum of such average values were noted and corresponding SDPs were chosen, as mentioned in the Table~\ref{table:suitability}. Further, suitability index was calculated by counting the SDP names across each application on both Minimum and Maximum of average columns and then percentage of their contribution, over all the metrics were calculated. Because, this helps to decide at what percent the SDP is suitable (Minimum of average column) or  not suitable (Maximum of average column). Finally, according to suitability index, the well suited SDP for each application was noted as mentioned in the Table~\ref{table:suitability}.

As mentioned earlier, Aeneas is BI application and according to suitability index, MQTT based SDP is well suited for this application with $43\%$, in-spite it has huge network consumption which is not acceptable for BI applications. However, it had good performance in other metrics.  The DFT is not suited due to higher processing time and disk utilization with $48\%$. In PocketSphinx application, MQTT based SDP has suitability index of $57\%$, where as it also has a highest not suitability index with $43\%$. The OSS had poor performance in all the aspects, but DFT has $0\%$ index for not suitability, this motivates to consider the DFT as a best suited SDP for PocketSphinx. Finally, for Video processing application, the performance of OSS was significant with suitability index of $51\%$ and $71\%$. 

\subsubsection{\textbf{Experience and future directions}}


The set of experiments and associated results in earlier subsections show that SDPs performance varies significantly according to end user application (CI, BI). An IoT has a stochastic and heterogeneity (latency intensive, CI, BI) nature of workloads. To process such  data oriented workloads the placement and design of the data pipeline mechanism on fog/cloud is quite necessary where our research fills this gap. 

However, while designing and implementing these SDP approaches significant portion of time was consumed in designing and developing the serverless functions in various proposed SDPs. The OSS consumed more time to design as it required more than five number of serverless functions as shown in Table~\ref{table:functions}, whereas DFT is least with maximum of three functions because of a set of built-in processors in Apache NiFi that could handle necessary utility operations such as \textit{PUTS3} to store a data in MinIO were used. But in MQTT based SDP and OSS we need to write them as functions. The DFT was best in designing and implementing, because state of art Apache NiFi data pipeline tool was used, which basically reduced efforts and easily integrated with serverless frameworks. \rone{The function templates and associated  Docker files of serverless functions including other utility source files are available in GitHub \footnote{https://github.com/shivupoojar/ServerlessDataPipelines}.}


\begin{table}\label{average}
\centering
\caption{Average performance metric values across three applications}
\footnotesize 
 \begin{tabular}{ || c c c c ||   }
 \hline
 Metric/Application & DFT & OSS & MQTT  \\ 
 \hline \hline
 Processing Time (m) & 20.97 & 23.97 & 21.77 \\
 CPU Utilization (\%) & 69 & 61 & 52 \\
 Memory Utilization (\%) & 97 & 85 & 86 \\
 Disk Read (KB) & 4 & 19 & 3 \\
 Disk Writes (KB) & 102 & 197 & 95 \\
 Network Receive (KB) & 15 & 31 & 33 \\
 Network Transmit (KB) & 22 & 43 & 63 \\ 
 \hline
 \end{tabular}
 \end{table} 

Apart from the design experience, the resource utilization metrics such as CPU, Memory, Disk Reads and Writes are important in serving the demands of IoT applications, because the resource demand from end-user requests vary with respect to the type of the application. For example, video application demands for maximum compute and bandwidth resources, whereas text processing application demands only for bandwidth.  So to investigate and analyze the resource utilization metrics and processing time, we calculated the average over all the three applications (on performance metrics) as shown in the  Table~\ref{average}.  

The DFT consumed highest CPU ($69\%$) and Memory ($97\%$), whereas least in network utilization ($15KB$, $22KB$) and processing time ($20.97m$), this indicates that DFT is best suitable for applications with huge bandwidth demand such as text processing. Further, OSS consumed moderate CPU ($61\%$), Memory ($85\%$) and network utilization ($31KB$, $43KB$), but had large number of disk read and writes ($19KB$, $197KB$). These \rone{results show} that OSS is best fit for video or image processing applications (bandwidth and compute intensive) due to lesser CPU, Memory and network utilization. On the other-side, MQTT-based SDP utilized the highest  network resource ($63KB$, $33KB$) and lesser CPU ($52\%$), disk and moderate memory resources ($86\%$). So, MQTT-based SDP is best suitable for compute intensive applications, however not suitable for bandwidth sensitive applications.

Even though the SDPs were designed and investigated significantly based on different performance metrics, certain challenges still exist and the key improvements can be undertaken. The future directions should focus on two aspects. Firstly, on how to minimize the processing time and resource utilization. In our analysis, synchronous mechanism was used to invoke the serverless functions leading to larger processing time and they should be tested with asynchronous mode. The serverless function scaling mechanism can also substantially reduce the processing time and using intelligent scaling mechanisms could overcome this issue. Secondly, how can SDPs be executed on fog/cloud in terms of dynamic and stochastic workloads? Serverless operations were focused to fog infrastructure in this work, however, to achieve user QoS expectations and to fulfill the dynamism nature, a part of pipeline could be dynamically executed in fog and rest in the cloud. So, several opportunities exist to implement such intelligent decision mechanisms.